\documentclass[10pt,aps,prb,twocolumn,amsmath,
			   notitlepage,floatfix,footinbib,superscriptaddress,showkeys]{revtex4-1}
\usepackage{graphicx}

\begin{document}

\title{Quantum Monte Carlo solution of the dynamical mean field equations in real time}

\author{Qiaoyuan Dong}
\author{Igor Krivenko}
\author{Joseph Kleinhenz}
\author{Andrey~E.~Antipov}
\affiliation{Department of Physics, University of Michigan, Ann Arbor, Michigan 48109, USA}
\author{Guy Cohen}
\affiliation{School of Chemistry, Tel Aviv University, Tel Aviv 69978, Israel}
\author{Emanuel Gull}
\affiliation{Department of Physics, University of Michigan, Ann Arbor, Michigan 48109, USA}

\date{\today}

\begin{abstract}
We present real-time inchworm quantum Monte Carlo results for single-site 
dynamical mean field theory on an infinite coordination number Bethe lattice. 
Our numerically exact results are obtained on the L-shaped Keldysh contour and, 
being evaluated in real-time, avoid the  analytic continuation issues typically 
encountered in Monte Carlo calculations. Our results show that inchworm Monte Carlo methods have now reached a state 
where they can be used as dynamical mean field impurity solvers and the 
dynamical sign problem can be overcome. As non-equilibrium problems can be 
simulated at the same cost, we envisage the main use of these methods 
as dynamical mean field solvers for time-dependent problems far from 
equilibrium.
\end{abstract}

\maketitle
\section{\label{sec:intro}Introduction}
The direct solution of quantum systems with many interacting degrees of freedom 
is believed to be intractable in general, and in order to understand the 
salient aspects of these systems, suitable approximations have to be employed. 
The dynamical mean field theory\cite{Georges96} is one such approximation. It 
is based on the realization that if the momentum dependence of the self-energy 
can be neglected, as occurs in certain infinite coordination number limits, the 
solution of a lattice model can be mapped onto that of an auxiliary quantum 
impurity model with self-consistently determined 
parameters.\cite{Georges92,Metzner89}

Quantum impurity models are amenable to numerical study, and the last decade 
has seen rapid advancement in the development of algorithms for their solution. 
In particular, remarkable progress was achieved by the continuous-time quantum 
Monte Carlo 
methods,\cite{Rubtsov2005,Werner2006,Gull2008,Otsuki2007,Prokofev2007,Prokofev2008}
which by now are the standard methods for studying 
multi-orbital and cluster impurity problems;\cite{Gull11_RMP} as well as in 
numerical renormalization group;\cite{Pruschke08} density matrix 
renormalization group;\cite{Wolf14} and configuration interaction 
methods.\cite{Knowles1984,Zgid12}

Modern quantum Monte Carlo methods are numerically exact, in the sense that 
they can provide results that converge to the exact answer with an uncertainty 
proportional to the square root of the number of stochastic samples taken. 
However, because these methods are formulated within an imaginary-time 
statistical mechanics formalism, real-frequency data such as spectral functions 
needs to be extracted from an ill-conditioned analytic continuation 
procedure,\cite{MaxentReview,Bryan1990,Vidberg1977,Sandvik1998,Beach2004,Mishchenko2000,
Fuchs2010,Prokofev2013,Otsuki2017}
in which these uncertainties are exponentially amplified.
Furthermore, studying 
systems in general nonequilibrium states or under time-dependent driving beyond 
linear response is not possible.

The desire to understand nonequilibrium transport in correlated impurity models 
motivated the development of real-time generalizations of continuous-time 
quantum Monte Carlo methods.\cite{Muhlbacher08,Schiro09,Werner10,Antipov16}
These early methods require exponentially increasing computer time as a 
function of the simulated time due to the dynamical sign problem, i.e. the 
sign problem occurring when real-time dynamics is evaluated. 
The development of bold-line impurity solver 
algorithms\cite{Gull10_bold,Gull11_bold,cohen_greens_2014,cohen_greens_2014-1} 
substantially alleviated this sign problem and increased the accessible 
parameter space. A recent development, the inchworm quantum Monte Carlo 
method,\cite{Cohen15} showed a reduction of the computational scaling from 
exponential to quadratic, effectively eliminating the dynamical sign problem 
altogether.

For dynamical mean field 
applications\cite{Eckstein2009Quench,Eckstein2010QuenchSpectrum,
Eckstein2011Bloch,Tsuji2011} of the inchworm method, two more components are 
necessary: the ability to obtain two-time response functions, 
and the extension of the method to an initial thermal ensemble. Both components 
have recently been implemented,\cite{Antipov17} allowing us to test the method. 
As a first application, we address a lattice model for which the dynamical mean 
field approximation becomes exact: the Hubbard model on the infinite 
coordination number Bethe lattice. While we see the main application of our 
method in non-equilibrium,\cite{NonEqDMFTReview} we demonstrate its power here 
for the equilibrium case, where a large variety of well-developed and 
competitive methods are available and the physics is well understood.

The paper is organized as follows. In Section~\ref{sec:methods}, we write down 
the lattice model, recapitulate the dynamical mean field theory, introduce our 
impurity solver, discuss how self-energies are extracted, and explain how linear 
prediction can be used to obtain spectral functions. In 
Section~\ref{sec:results} we present our results, including real-time Green's 
functions, real-time self-energies and real-frequency spectral functions with 
and without linear prediction. Finally, in Section~\ref{sec:conclusions}, we 
discuss our conclusions and outlook. The appendices contain further technical 
and numerical information.

\section{\label{sec:methods}Methods}
\subsection{Lattice model and dynamical mean field theory}
We study the repulsive Hubbard model in equilibrium on the infinite coordination number Bethe lattice. The Hamiltonian is given by
\begin{align}\label{eq:Ham}
	\hat H = -\tilde v\sum_{\langle ij\rangle} c^\dagger_{i\sigma} c_{j\sigma} +
	U\sum_i n_{i\uparrow} n_{i\downarrow},
\end{align}
where $\tilde v$ denotes the hopping matrix element, and $\sigma$ denotes the 
spin index. The operators $c^\dagger_{i\sigma}$/$c_{i\sigma}$
create/annihilate electrons with the spin $\sigma$ on the corresponding 
site $i$. $U$ is the on-site Coulomb repulsion between electrons with opposite 
spins.
We restrict our discussion to the paramagnetic solution of a half-filled 
infinite coordination number Bethe lattice. The hopping matrix element must be
properly scaled with the coordination number $Z$ to remain finite in the 
limiting case, $\lim_{Z\to\infty}(\tilde v \sqrt{Z}) = v$.\cite{Metzner89}
Throughout this paper we employ dimensionless units by dividing/multiplying all 
energy/time scales by the hopping strength $v$. The Bethe lattice is
characterized by a semi-elliptical noninteracting density of states $D(\epsilon) = 
\sqrt{4 - \epsilon^2} / (2\pi)$. The model's interacting self-energy 
$\Sigma_{i\sigma,j\sigma'}(t,t')$, corresponding to Eq.~\ref{eq:Ham}, is purely 
local (\emph{i.e.} zero for $i \neq j$), and its local part 
$\Sigma_{i\sigma,i\sigma'}(t,t')$ is equal to the self-energy of an auxiliary 
impurity model. The lattice Green's function can easily be obtained from this 
self-energy.\cite{Georges96}

We simulate the real-time dynamics of the model in equilibrium by means of the real-time dynamical mean field formulated on the $L$-shaped Keldysh contour.\cite{Monien2002,Freericks2006}
We define the impurity Green's function as the contour-ordered expectation value \cite{NonEqDMFTReview}
\begin{align}
G_\sigma(t,t') \equiv -i \left\langle \mathcal{T_C} c_\sigma(t) c_\sigma^{\dagger}(t') \right\rangle,
\end{align}
where $c_\sigma$ and $c_\sigma^\dagger$ denote impurity operators, such that the retarded Green's function is given by
\begin{align}\label{eq:retG}
G_\sigma^\mathrm{ret}(t,t') = -i \theta(t-t')\left\langle \left[c_\sigma(t),c_\sigma^\dagger(t')\right]_+  \right\rangle.
\end{align}

Time-translation invariance of the system in equilibrium implies that the two-time Green's function 
 only depends on time differences: $G_\sigma(t,t')=G_\sigma(t-t')$. The DMFT self-consistency condition for the Bethe lattice thus reads
\begin{align}
	\Delta_\sigma(t - t') = G_\sigma(t-t')
\end{align}
for any two times $t,t'$ on the L-shaped contour.

We solve the dynamical mean field equations iteratively, starting from a metallic, insulating, or high-temperature guess for the Green's function.

\subsection{Inchworm impurity solver}
Each DMFT iteration requires the solution of an impurity problem. We employ the inchworm quantum Monte Carlo solver\cite{Cohen15,Antipov17} to obtain a numerically exact solution of the Green's function of this single impurity Anderson model, up to some finite maximal time.

The inchworm method used here is a continuous-time (or diagrammatic) Monte Carlo method\cite{Gull11_RMP} based on sampling Feynman diagrams from a hybridization expansion. Unlike standard continuous-time methods,
the inchworm method takes advantage of the causal structure of propagation: information from easy-to-evaluate propagation over short times is reused to more efficiently evaluate propagation over longer times. The method has been shown to circumvent the dynamical sign problem in several models, diagrammatic expansions, and physical regimes.\cite{Cohen15,Antipov17,Chen17A,Chen17B} Its total run time (when applied to the hybridization expansion) grows quadratically with the simulated time. In practice, for the model and initial condition treated in the present work, times about an order of magnitude larger than possible with a bare algorithm can be accessed. Details of the solver implementation, and in particular of the Keldysh Green's function measurement, are reported in Ref.~\onlinecite{Antipov17}.

\subsection{Real-time self-energies}
\label{sec:sigma}

The local self energy function is of special interest in the DMFT context
as it contains all relevant information about single-particle correlations in the system.
Its retarded component $\Sigma^\mathrm{ret}(t-t')$ is defined as the solution of the Dyson equation (spin indices are omitted for clarity)
\begin{multline}\label{dyson}
	G^\mathrm{ret}(t-t') = G_0^\mathrm{ret}(t-t') +\\+
	\iint_{0}^{t_\mathrm{max}} dt_1 dt_2 G_0^\mathrm{ret}(t-t_1) \Sigma^\mathrm{ret}(t_1-t_2) 
	G^\mathrm{ret}(t_2-t').
\end{multline}
Here, $G_0^\mathrm{ret}(t-t')$ is the retarded Green's function of the 
noninteracting lattice, $G^\mathrm{ret}(t-t')$ is obtained numerically as 
a result of a DMFT/inchworm calculation, and $t_\mathrm{max}$ is a maximum simulation time.

Eq.~\eqref{dyson} is a Volterra integral equation of the first kind with respect to $\Sigma^\mathrm{ret}$.
In a numerical implementation, $G^\mathrm{ret}_0(t)$ and $G^\mathrm{ret}(t)$ are known on some finite time mesh (for example, a uniform grid). It is therefore natural to project the integral equation onto the mesh, and use numerical linear algebra methods to solve the resulting linear system.
In principle, the solution of Eq.~(\ref{dyson}) could be obtained in Fourier space by applying the convolution theorem. However, this approach is not practical here, as $G^\mathrm{ret}(t)$ is known only up to a finite time, and does not generalize to the non-equilibrium case without additional modifications.

Given a number of time slices $N_t$, we introduce mesh nodes $t_i = i\Delta t$ with
$i\in\left\{ 0,1,\ldots,N_{t}-1\right\}$ and $\Delta t = t_\mathrm{max} / (N_t - 1)$.
A discretized version of the Dyson equation then reads
\begin{multline}\label{dyson_discretized}
	G^\mathrm{ret}(t_i-t_j) = G_0^\mathrm{ret}(t_i-t_j) +\\+
	(\Delta t)^2 \sum_{k,l=0}^{N_t-1} G_0^\mathrm{ret}(t_i-t_k) 
	\Sigma^\mathrm{ret}(t_k-t_l) G^\mathrm{ret}(t_l-t_j) w_{kl}.
\end{multline}
The quadrature weights $w_{kl}$ define the integration method, and at this point we do not specify their precise form. Using matrix notation $F_{ij} = F(t_i - t_j)$ we find
\begin{equation}\label{dyson_solution}
	\Sigma^\mathrm{ret}_{ij} w_{ij} =
	\frac{(G_0^\mathrm{ret})^{-1}_{ij} - (G^\mathrm{ret})^{-1}_{ij}}
	{(\Delta t)^2}.
\end{equation}
The numerical matrix 
inversion used here is stable, as all retarded Green's functions are 
represented by lower triangular matrices with $G^\mathrm{ret}(0^+) = -i$ on the 
main diagonal, such that their condition number is 1. The $(\Delta t)^2$ in the 
denominator of (\ref{dyson_solution}) suggests that this procedure is similar to
numerical calculation of the second derivative. The number of time slices at 
which $G^\mathrm{ret}(t_i)$ is known is limited by the computational 
effort required by the inchworm QMC algorithm. In order to make the numerical 
differentiation accurate, we choose $N_t$ much larger than used in the inchworm
simulation and employ a cubic spline interpolation of $G^\mathrm{ret}$ between
the original nodes.

Only certain choices of quadrature weights give a convergent solution in the small $\Delta t$ limit.
In his study of the one-dimensional Volterra equations of the first kind,
\textcite{Linz1969} showed that trapezoidal and rectangular rules
are convergent, whereas higher order quadrature methods in general are not.
Using this result as a starting point, we construct $w_{ij}$ as possible
direct products $w_i w_j$, where $w_i$ and $w_j$ correspond to the rectangular rule
with the first/last point excluded and to the trapezoidal rule. Most combinations
can be ruled out, as they cause $\Sigma^\mathrm{ret}(t_\mathrm{max})$ to diverge in the
small $\Delta t$ limit. We found that the following choice:
\begin{equation}\label{quadrature_weights}
	w_{i\geq j} = \left\{
	\begin{array}{ll}
		0,& i = 0,\\
		0,& j = N_t-1,\\
		0,& i = j,\\
		1/2,& j = i-1,\\
		1,& \mathrm{otherwise}.
	\end{array}
	\right.
\end{equation}
yields stable and accurate results.
The first two lines in (\ref{quadrature_weights}) show that we choose a
rectangle rule approximation excluding the first slice from the $t_1$-integral 
and the last slice from the $t_2$-integral. Values on the first sub-diagonal
receive half the weight, because retarded functions are proportional to
$\theta(t-t')$. Finally, $w$ as well as all other matrices entering the equation
must be of Toeplitz form in equilibrium. We therefore also set the main 
diagonal of $w$ to zero.

Equations (\ref{dyson_solution}) and (\ref{quadrature_weights}) allow for extracting
all mesh values of the self energy, except $\Sigma^\mathrm{ret}_{ii} = 
\Sigma^\mathrm{ret}(0^+)$. This element is known analytically,
see Appendix \ref{app:short_time}.

\subsection{Linear Prediction}\label{sec:linpred}
Our simulations are performed up to a finite maximum real time
$t_{\rm max}$ determined by the available computational resources. However, many
quantities are best described in the real-frequency domain, and are therefore
expressed as Fourier transforms over the entire time axis. When Fourier transforming
quantities with a hard time cutoff, the transform can be expressed as a convolution
with a sampling kernel proportional to
$\text{sinc}(\omega)=\frac{\sin(\omega)}{\omega}$ in the frequency domain, i.e.
\begin{align}
  \tilde{A}(\omega) = \int_{-\infty}^{\infty} d\omega' K(\omega, \omega') A(\omega'),
\end{align}
where $\tilde{A}(\omega)$ is the result from a Fourier transform with data up to finite times,
$A(\omega)$ is the true result and the convolution kernel is given by
\begin{align}
  K(\omega, \omega') = \frac{t_{\rm max}}{2\pi} {\rm sinc}\left(\frac{\left(\omega - \omega'\right) t_{\rm max}}{2}\right).
\end{align}
This convolution introduces broadening and unphysical oscillations into the
spectral function. Linear prediction is a technique to remove these artifacts
by using a physically motivated extrapolation scheme to extend data beyond the
maximum simulated time. Linear prediction has previously been used for this
purpose within the framework of $t$-DMRG\cite{White08, Barthel09, Pereira12,
Ren12}. We start with the ansatz that the value of the signal (in this case the Green's function)
at the $n$-th real time point is a linear function of its previous $p$ values, i.e.
\begin{align}
  \tilde{x}_n = - \sum_{i = 1}^p a_i x_{n - i}.
\end{align}
It can be shown that this corresponds to fitting the function in time to a superposition of $p$ complex
exponential terms. In the case of a Green's function dominated by a few
isolated poles, this approximation is justified. For the data presented here, the validity needs to be assessed by systematically varying $p$.

In order to use this model for extrapolation, the coefficients $a_i$ must be fit
to the known data. This is done over the region 
$\left(t_{\rm max} - t_{\rm fit}, t_{\rm max}\right)$ in
order to exclude spurious short time behavior from the fit results.
The linear prediction ansatz leads to a matrix equation
\begin{align}
  Q \boldsymbol{a} = -\boldsymbol{x},
\end{align}
where $Q_{nk} = x_{n - k}$ is an $N \times p$ complex matrix with $N$ the number of points in $\left(t_{\rm max} - t_{\rm fit}, t_{\rm max}\right)$.
Solving this in the least squares sense leads to the normal equations
\begin{align}
  R \boldsymbol{a} = - \boldsymbol{r}.
\end{align}
This is written in terms of the autocorrelations of the data,
\begin{align}
  R_{ji} = \sum_{n} x^*_{n - j} x_{n - i}; &&
  r_j    = \sum_{n} x^*_{n - j} x_n.
\end{align}
These matrix equations are often unstable and require some form of
regularization. Here we choose the simple regularization, 
$R^{-1} \to \left(R + \epsilon I\right)^{-1}$ 
and check that the results are not strongly affected by the regularization
parameter.\cite{Barthel09} With this scheme, the coefficients $a_i$
are readily obtained and the Green's function can be extrapolated until it
decays to zero.  The procedure yields spectral functions free from unphysical
finite time oscillations, at the cost of some additional systematic uncertainty
due to the assumptions imposed by the linear prediction ansatz.  We emphasize that
 linear prediction is only applied to our converged data as a post-processing routine,
in order to interpret the real-time results as functions of frequency. The DMFT
/ inchworm iteration procedure preceding it is formulated in terms of real, finite times
only, and independent of the linear prediction formalism.

\begin{figure}[tbh]
\includegraphics[width=\columnwidth]{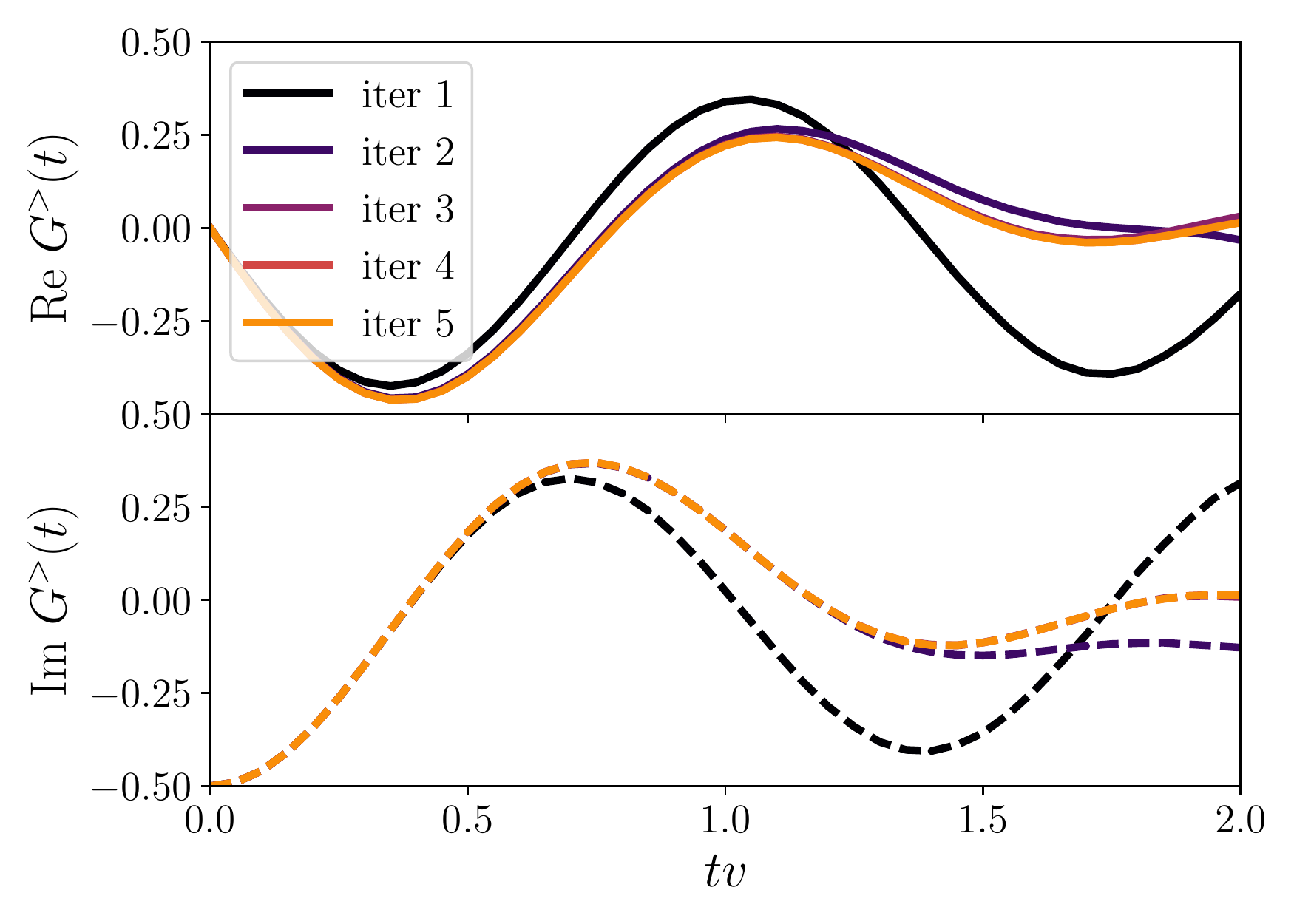}
\caption{
Real and imaginary parts of the greater dynamical mean field Green's function $G^>(t)$ as a function of real time $t$ up to a maximum time of $tv=2$. $U/v= 4$, $T/v= 0.05$, equilibrium. Shown is the convergence with DMFT iteration.
}
\label{fig:dmft_Gt_convergence}
\end{figure}
\section{\label{sec:results}Results}
Our impurity solver obtains results for $G(t,t')$ as a function of real times. A typical raw simulation output is shown in Fig.~\ref{fig:dmft_Gt_convergence}, 
which shows the real (solid lines) and the imaginary part (dashed lines) of the greater Green's function up to a maximum time $tv=2$ and at a low temperature of $T/v=0.02$. The interaction strength is taken to be $U/v=4$, equal to the full bandwidth of the infinite coordination number Bethe lattice.
The error bars of the measured Green's function could, in principle, be estimated as standard deviation from a set of completely independent DMFT/inchworm runs. With our present implementation this approach has proven too computationally expensive.

The inchworm Monte Carlo method is only exact when two numerical parameters are controlled. The first of the parameters is the discretization of the imaginary and the real time branch $\Delta t$, which we chose to be $\Delta tv=0.05$. Second, the maximum order to which diagrams in the inchworm expansion are considered. We find, especially in the metallic low temperature regime, that results converge at an inchworm expansion order of around seven. As the inchworm order is directly related to the number of crossings considered in an $N$-crossing approximation,\cite{Cohen15} this result implies that results from non-crossing or one-crossing approximations are not valid in this parameter regime. Throughout this paper, all the results presented are converged in both $\Delta t$ and maximum order.

We observe that the dynamical mean field solution converges in a causal manner, in the sense that results are converged within one iteration by time $tv=0.3$, within two iterations by time $tv=1.0$, and results up to $tv=2$ are indistinguishable between iterations $4$ and $5$, indicating that the self-consistency loop has converged. While this causal convergence can be used to avoid the usual dynamical mean field iteration scheme and replace it with a direct solution of the self-consistency equations,\cite{NonEqDMFTReview} this scheme has not been employed here.

\begin{figure}[htb]
\includegraphics[width=\columnwidth]{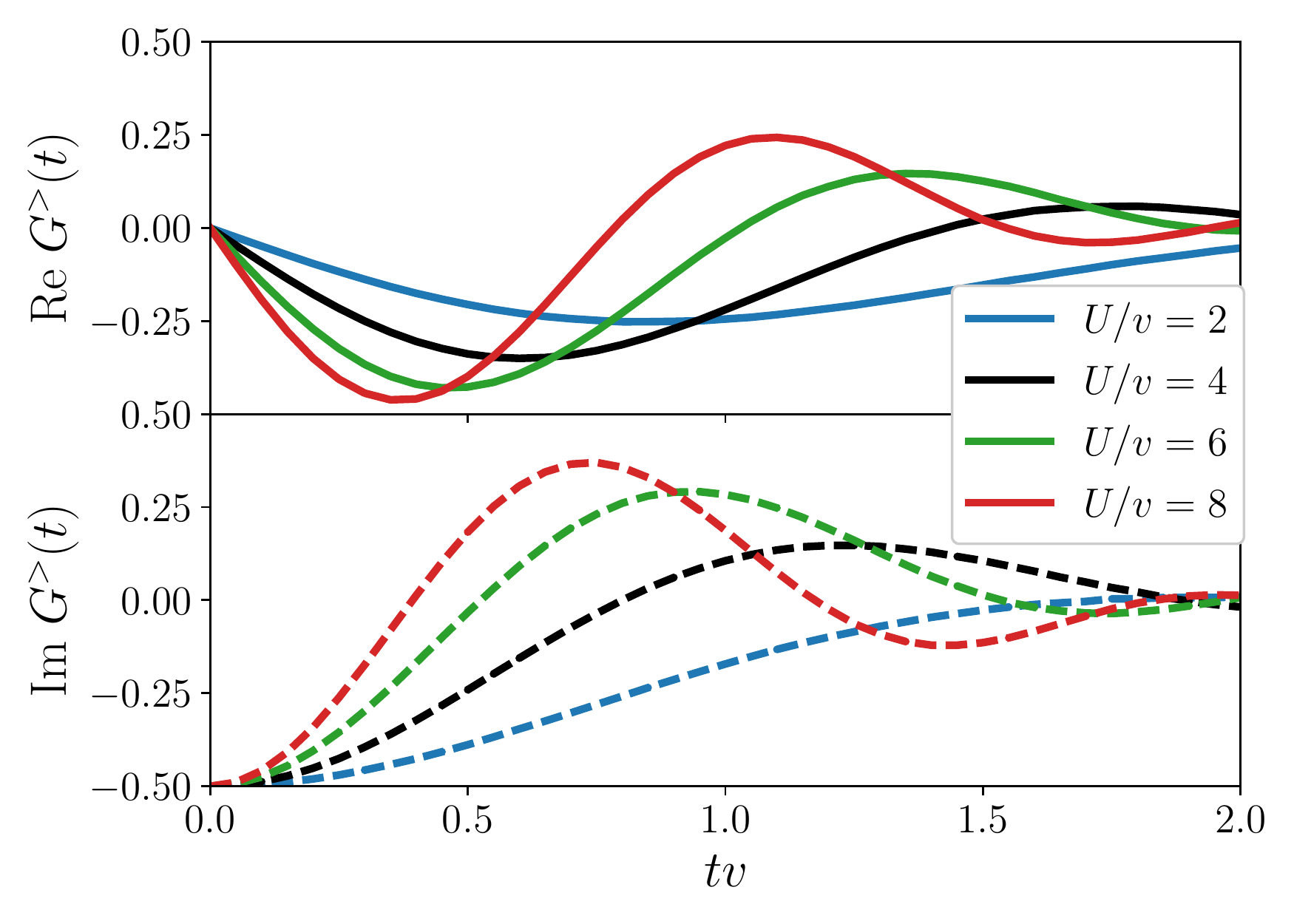}
\caption{
Real and imaginary parts of the greater DMFT Green's function $G^>(t)$, for times up to $tv=2$, with $T/v=0.5$, half filling, equilibrium, at  on-site interaction strengths $U/v\in\{2.0,4.0,6.0,8.0\}$.
}
\label{fig:DMFT_Gt_diffU}
\end{figure}
As the model is tuned from metallic to insulating, the shape of the real-time Green's functions changes substantially, from slow oscillations at weak coupling to rather rapid, quickly decaying oscillations at large interaction strength. This is shown in Fig.~\ref{fig:DMFT_Gt_diffU}, for interaction strengths $U=2/v$, $U/v=4$, $U/v=6$, and $U/v=8$, at a relatively high temperature of $T/v=0.5.$ 

\begin{figure}[htb]
\includegraphics[width=\columnwidth]{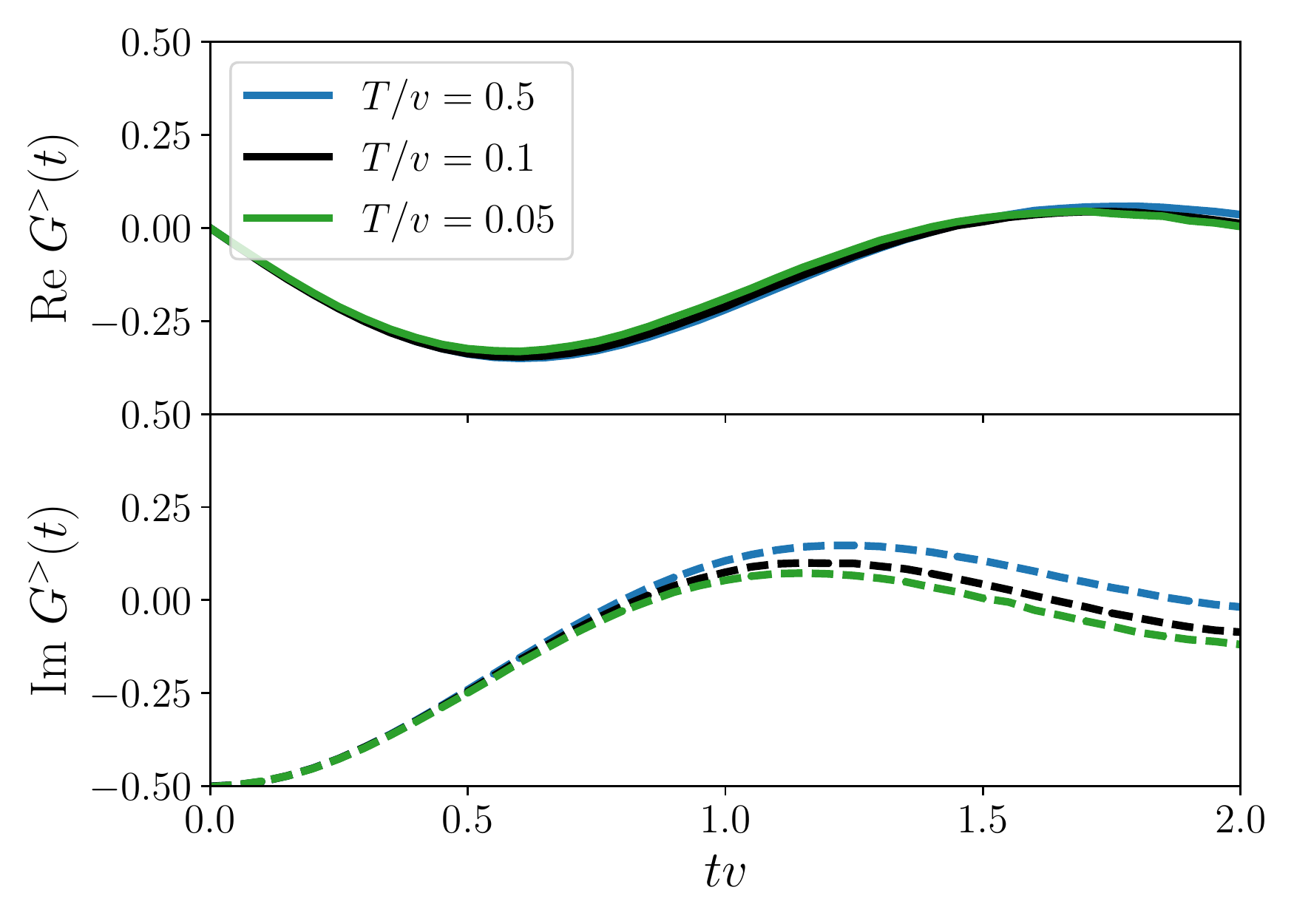}
\caption{
Real and imaginary parts of the greater DMFT Green's function $G^>(t)$, for times up to $tv=2$, half filling, equilibrium, for $U/v = 4$, at temperatures $T/v\in\{0.5,0.1,0.05\}$. 
}
\label{fig:DMFT_Gt_diffT}
\end{figure}
In contrast, lowering the temperature by a factor of $10$, as shown in Fig.~\ref{fig:DMFT_Gt_diffT}, causes relatively little change in the Green's function, with both oscillation frequency and amplitude staying more-or-less invariant for the time range simulated. Fig.~\ref{fig:DMFT_Gt_diffT} shows a regime in which a quasi-particle peak emerges upon cooling. However, the properties of this peak are not obvious in the data for the greater Green's function shown here.

\begin{figure}[htb]
\includegraphics[width=\columnwidth]{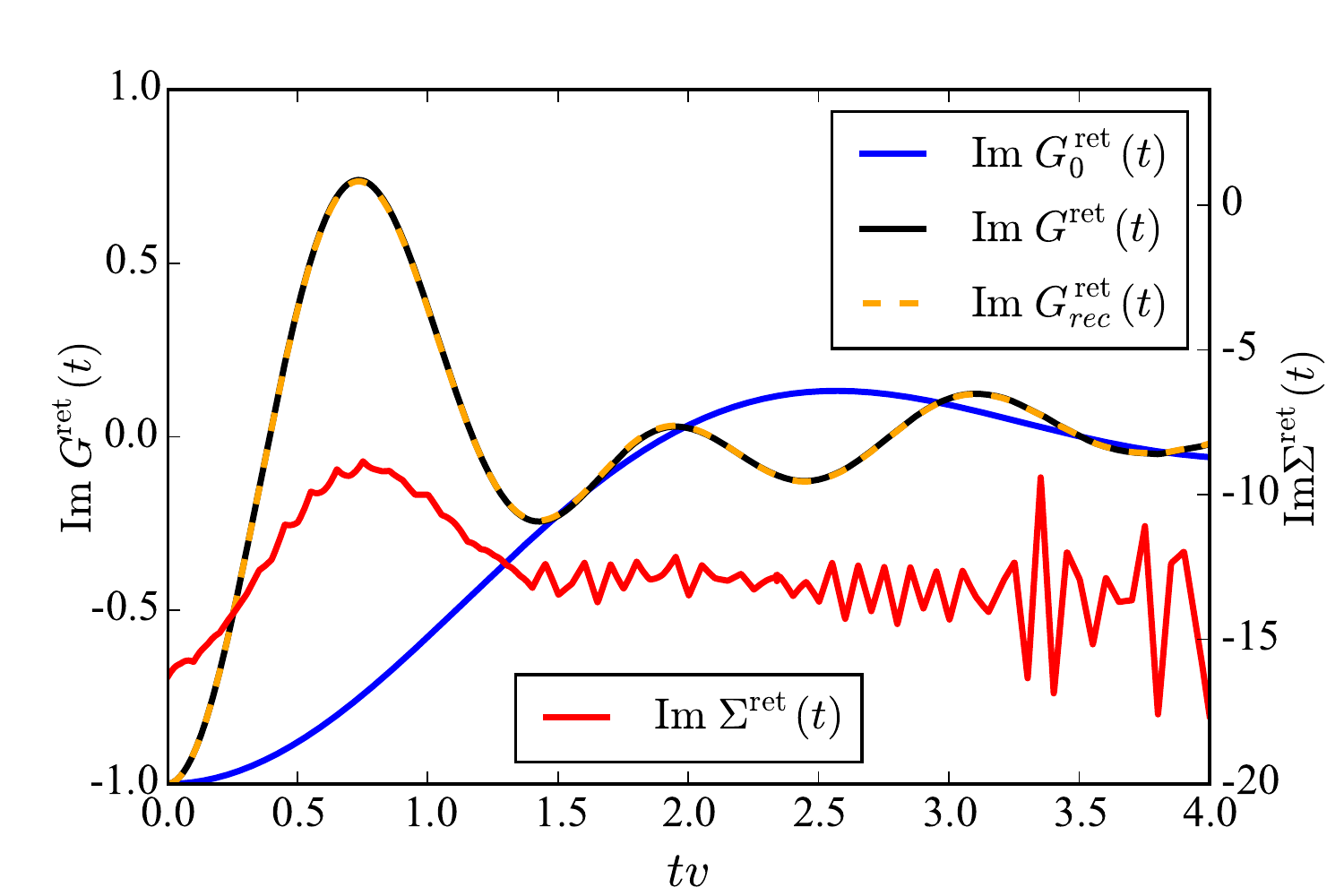}
    \caption{
		Retarded components of the DMFT Green's function, bare Green's function and
        self-energy computed for $U/v = 8.0$ at half-filling and temperature $T/v = 0.5$.
        $G_{rec}^\mathrm{ret}(t)$ (dashed orange curve lying on top of the black one) is a
        Green's function reconstructed by iterative substitution of $\Sigma^\mathrm{ret}(t)$
        into the Dyson equation. Data obtained using 2001 interpolation slices.
    \label{fig:Sigma_t_U8}}
\end{figure}
Using the procedure described in Sec.~\ref{sec:sigma}, we can directly extract a real-time self-energy. Fig.~\ref{fig:Sigma_t_U8} shows the imaginary parts of the noninteracting and the interacting retarded Green's functions (left vertical axis)  along with the imaginary parts of the computed self-energy (right vertical axis) as a function of time. The orange curve is the Green's function reconstructed from replacing the bare Green's function and the self-energy into eq.~\eqref{dyson_discretized}, showing that the scheme for extracting the real-time self-energy, which is numerically demanding, has converged. The numerical noise visible in the real-time self-energy may be used to qualitatively assess the size of the Monte Carlo errors intrinsic to this simulation. However, we note that in the case of inchworm simulations this approach typically underestimates the errors, and a more complicated procedure is required if one is interested in rigorous error estimates.\cite{Cohen15,Antipov17}

\begin{figure}[htb]
\includegraphics[width=\columnwidth]{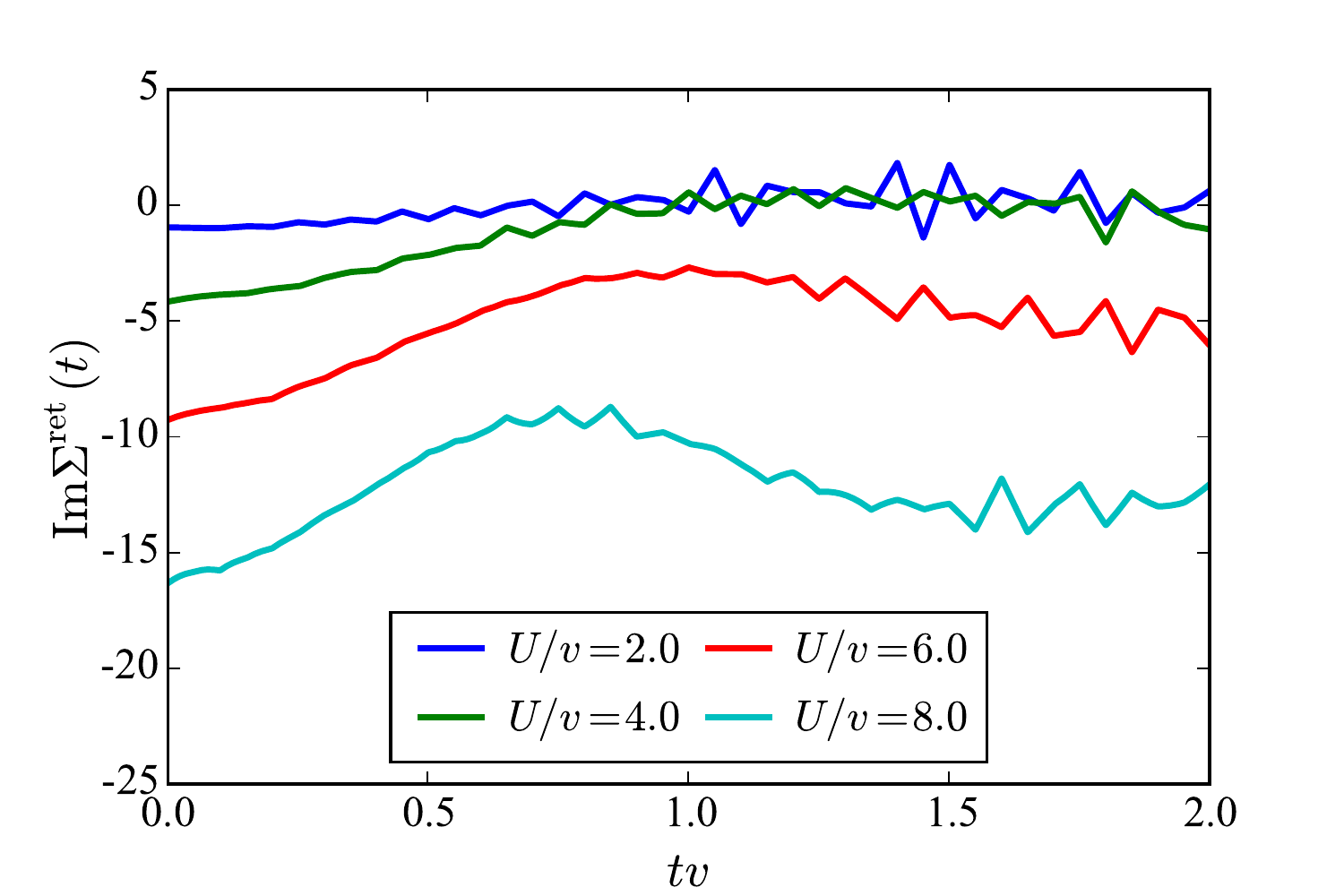}
    \caption{
		Retarded component of the imaginary part of the DMFT self-energy for 
		interaction strengths $U/v\in\{2.0,4.0,6.0,8.0\}$ at half-filling
        and temperature $T/v = 0.5$.
    \label{fig:Sigma_t_Us}
}\end{figure}
Fig.~\ref{fig:Sigma_t_Us} shows the evolution of the imaginary part of the real-time self-energy with interaction strength, for the same parameters as shown in Fig.~\ref{fig:DMFT_Gt_diffU}. The self-energy decays to zero within the accessible times in the weak coupling limit.

In contrast, the insulating regime shows a self-energy consistent with a constant in the long-time limit. This behavior is caused by the DMFT mechanism for opening a gap, which requires a pole at zero frequency (at half-filling) in the self-energy, such that $\text{Im}\Sigma(\omega) = \Delta^2\delta(\omega)+\text{Im}\Sigma^\text{reg}(\omega)$, where $\Delta$ denotes the Mott half-gap size and $\Sigma^\text{reg}$ the non-divergent part of the self-energy. In the atomic limit, one would expect $\Delta \sim U/2$. For the lattice problem with non-zero hybridization, the gap sizes are smaller.

\begin{figure}[htb]
\includegraphics[width=\columnwidth]{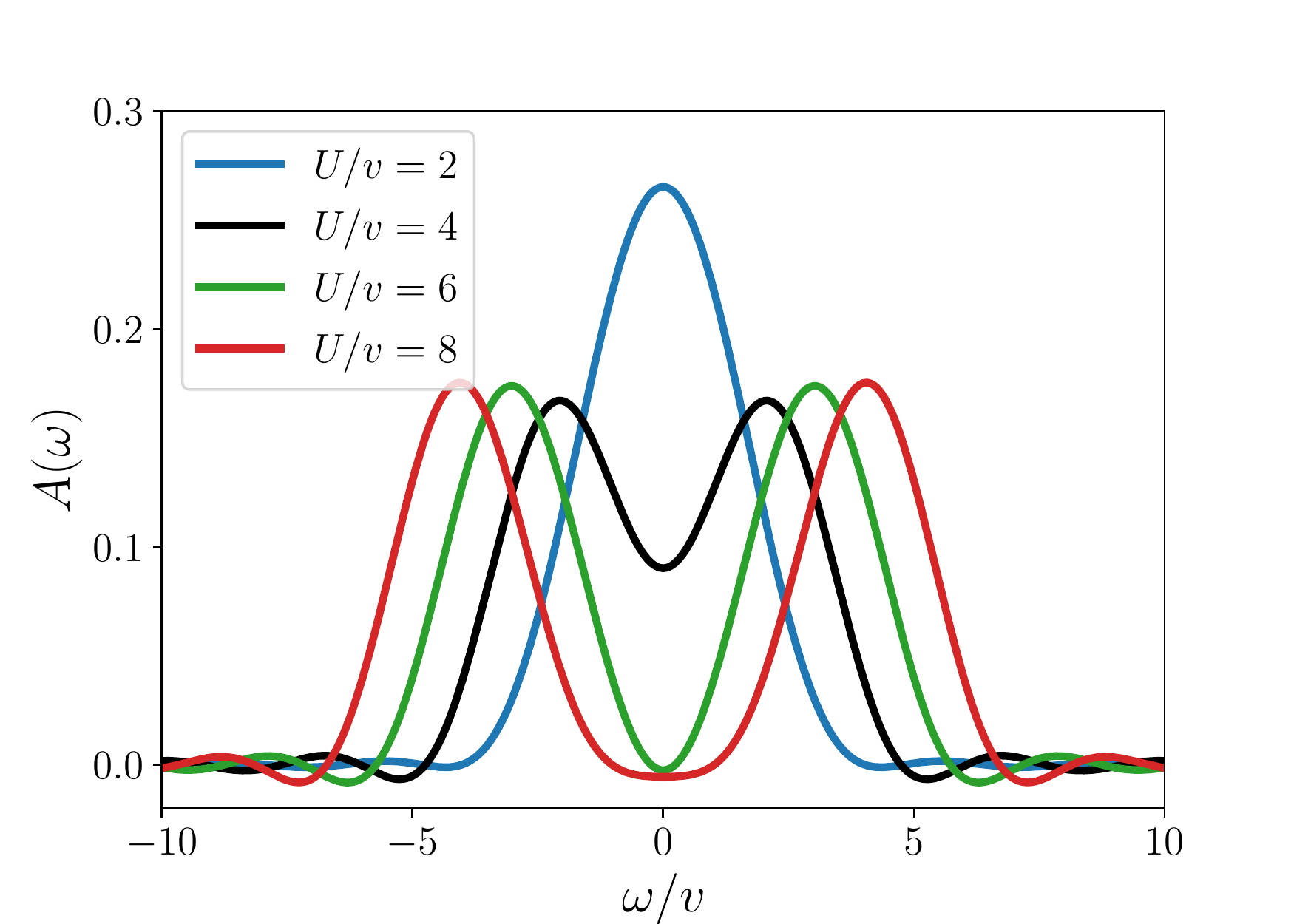}
\caption{
The converged DMFT spectral function $A(\omega)$ obtained by directly 
performing the Fourier transform on the real time Green's function with a 
cutoff at $t_{\text{max}}v = 2$ with $T/v=0.5$ and at on-site interaction strengths
$U/v\in\{2.0;4.0;6.0;8.0\}$.
}
\label{fig:DMFT_Aw_diffU}
\end{figure}
Interacting spectral functions in real frequency, 
$A(\omega)=-\frac{1}{\pi}\text{Im}G^\mathrm{ret}(\omega)$, are of principal 
interest in dynamical mean field calculations because they allow for direct 
comparison with photoemission experiments. Obtaining $A(\omega)$ in 
imaginary-time formulations requires an ill-conditioned analytical continuation 
procedure, such as the maximum entropy algorithm. The real-time formulation 
avoids this. However, the fact that the real-time Green's function is only 
known up to a finite maximum time implies that spectral functions can only be 
resolved with a  frequency resolution proportional to the inverse of that maximum time. Fig.~\ref{fig:DMFT_Aw_diffU} shows spectral functions extracted from the 
data in Fig.~\ref{fig:DMFT_Gt_diffU} and corresponding to the self-energies in 
Fig.~\ref{fig:Sigma_t_Us}. It demonstrates the metal-insulator crossover as the on-site interaction strength increases.
Due to the finite frequency resolution, sharp features are absent and 
part of the spectral function turns negative, especially in the higher 
frequency regions.

\begin{figure}[htb]
  \includegraphics[width=\columnwidth]{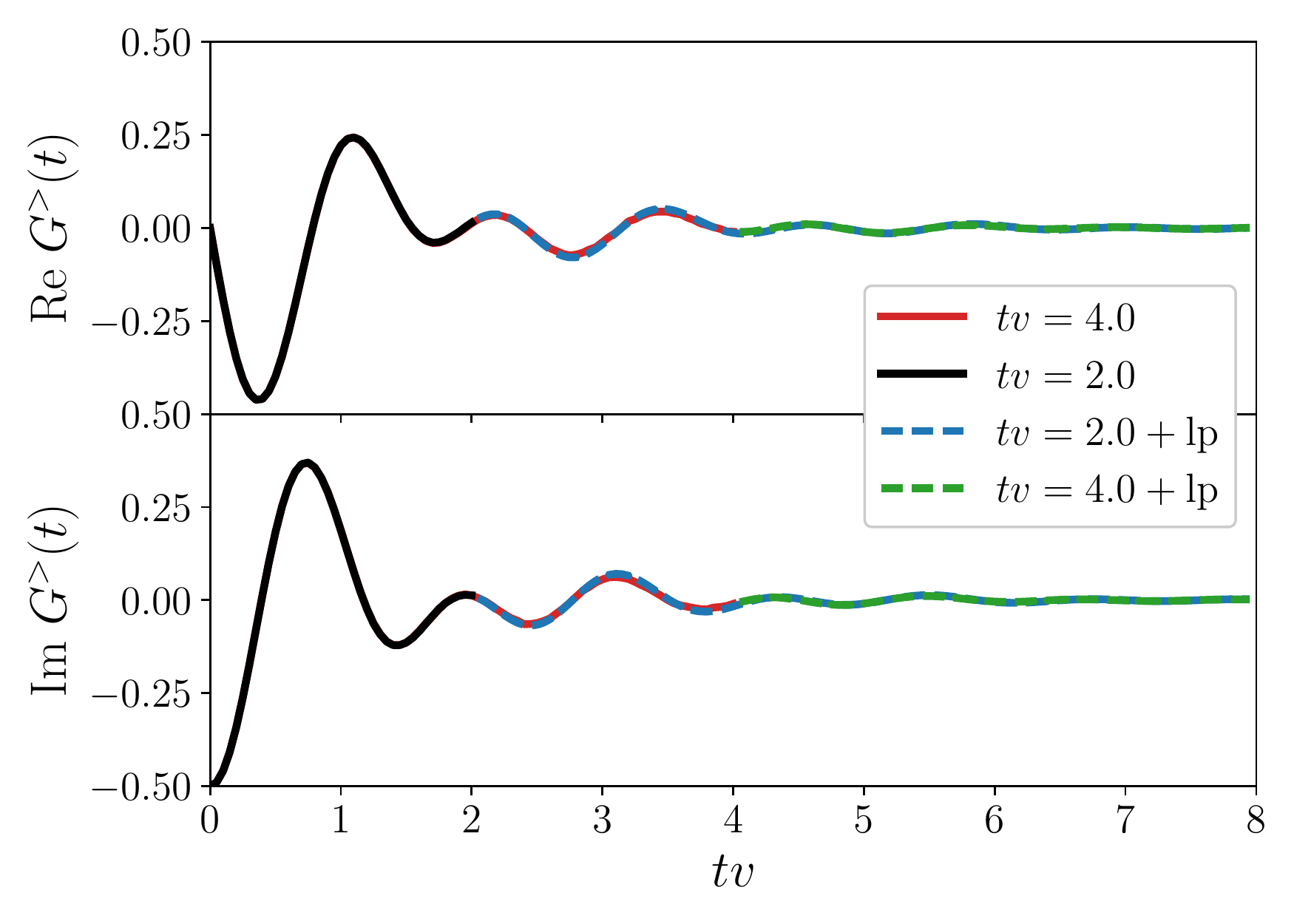}
  \caption{Comparison of raw Green's function data up to $tv=2$ (black) and
  $tv=4$ (red) with results from linear prediction of the $tv=2$ data (blue) and
  $tv = 4$ data (green) for $U/v = 8$, $T/v = 0.5$. For the linear prediction we 
  use $p = 9$, $t_{\rm fit}v = 1.0$.}
  \label{fig:gf_lp_validation}
\end{figure}
The linear prediction method described in Sec.~\ref{sec:linpred} is an interpolation routine designed to replace the sharp cutoff of $G(t)$ at the maximum time $t_\text{max}$ with a smoothly decaying function corresponding to a set of poles in the complex plane. As is evident in Fig.~\ref{fig:gf_lp_validation}, data obtained for times up to $tv=2$ and extrapolated up to $tv=4$ using linear prediction approximates well our data directly computed by running the dynamical mean field algorithm up to $tv=4$.

\begin{figure}[htb]
  \includegraphics[width=\columnwidth]{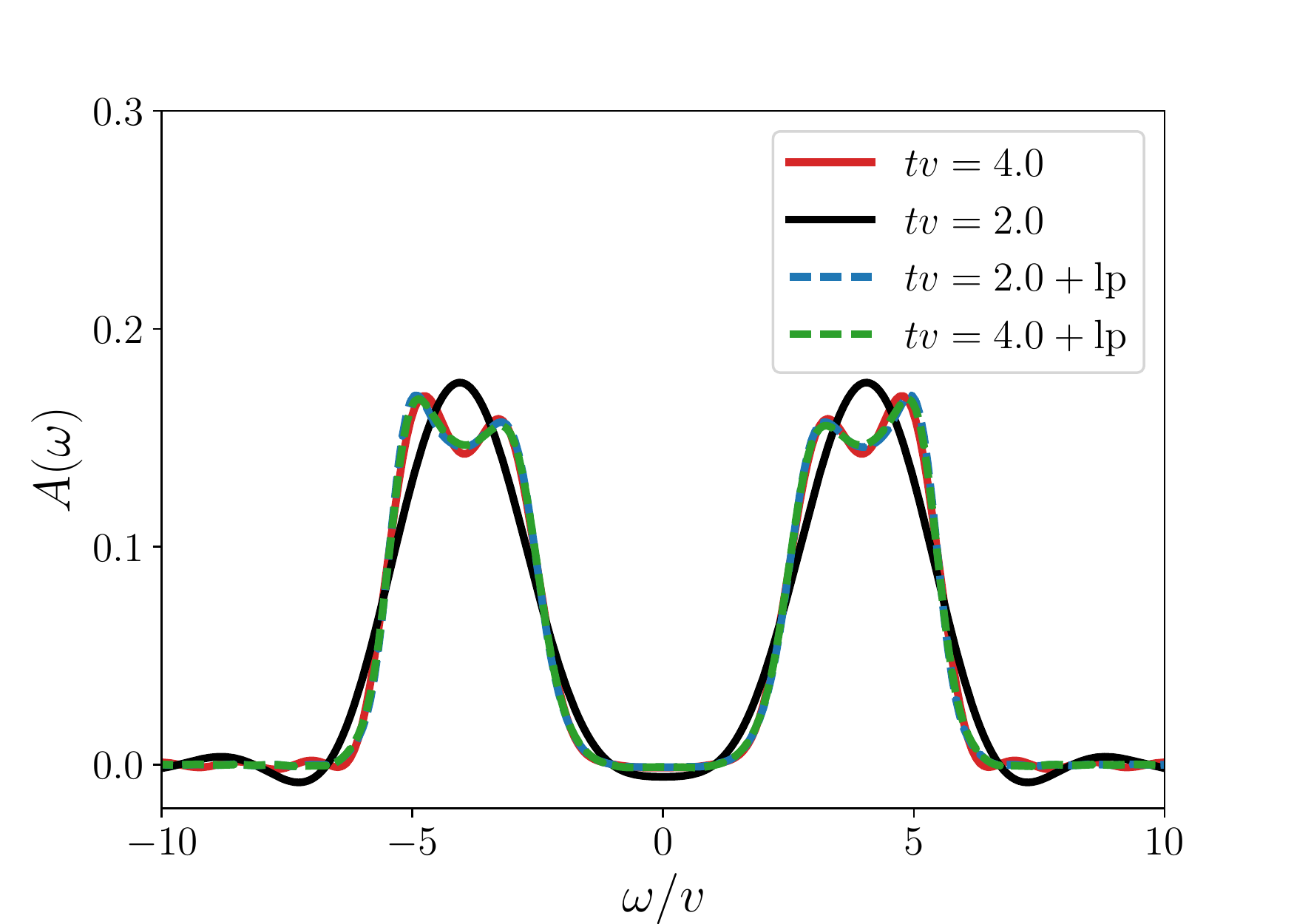}
  \caption{Comparison of spectral function obtained from raw Green's function data up to $tv = 2$ (black) and $tv = 4$ (red) with results from linear prediction of the $tv = 2$ data (blue) and $tv = 4$ data (green) for $U/v = 8, T/v = 0.5$. For the linear prediction we use $p = 9$, $t_{\rm fit}v = 1.0$.}
  \label{fig:spectrum_lp_validation}
\end{figure}
The linear prediction results can then be used to obtain spectral functions that do not suffer from the `ringing' phenomenon. Fig.~\ref{fig:spectrum_lp_validation} shows the results of this procedure in practice: while the straightforward continuation of the data up to time $tv=2$ shows only a coarse frequency resolution and has substantial negative contributions between frequencies of $\omega/v=5$ and $10$, the corresponding extrapolated data shows a double-peak feature and is positive. Longer-time data (not based on linear prediction) corroborates the double-peak structure and the slightly larger gap, and similarly does not exhibit a negative contribution to the spectral function.

Linear prediction is an extrapolation procedure that needs to be carefully 
controlled. Results should not depend on the choice of cutoff time or the 
number of poles interpolated. These concerns are addressed in Appendix
\ref{app:linpred_robust}.

\begin{figure}[htb]
\includegraphics[width=\columnwidth]{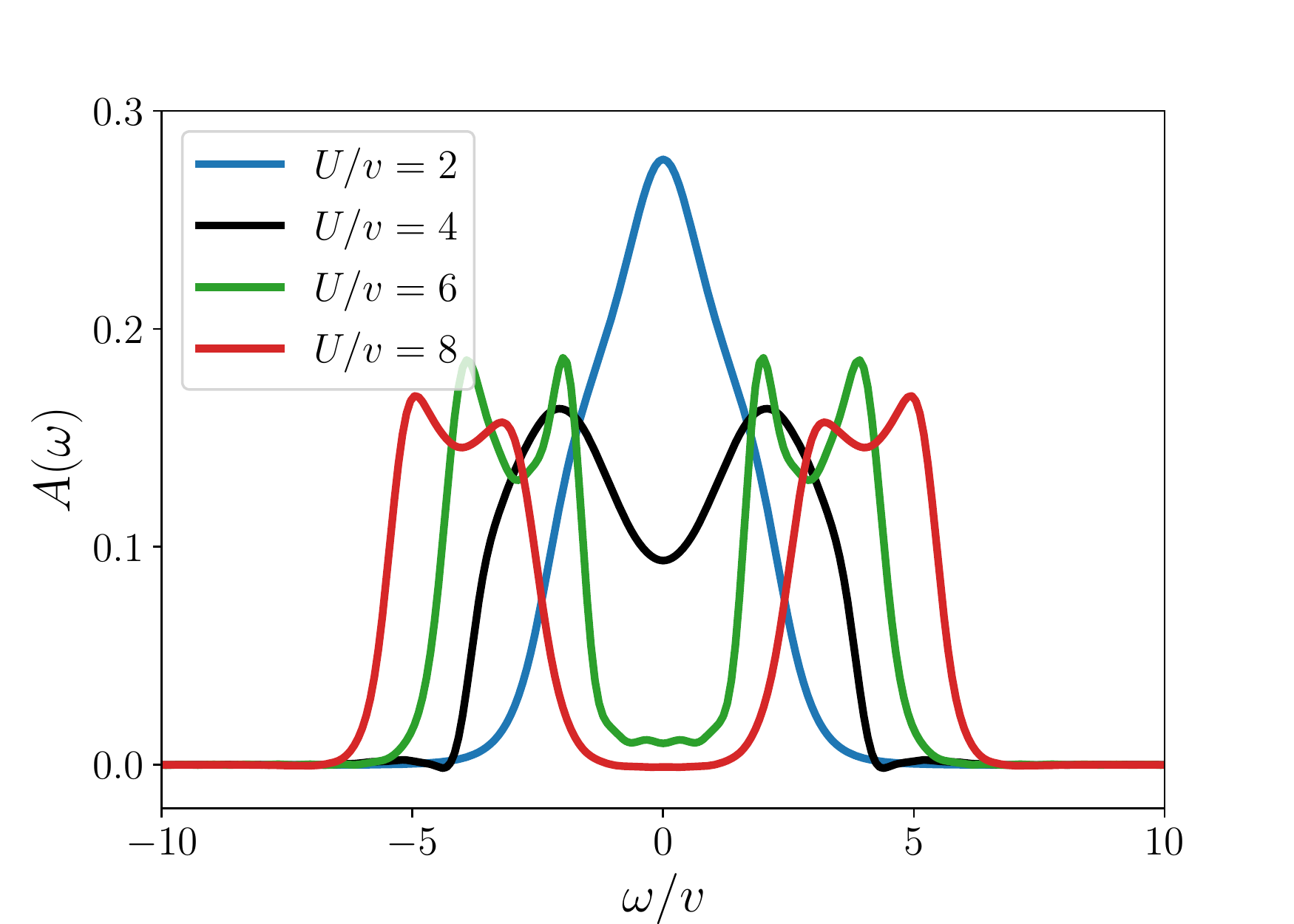}
\caption{
  The converged DMFT spectral function $A(\omega)$ obtained by extrapolating
  the real-time Green's function from $tv = 2.0$ to $tv = 10.0$ using linear
  prediction with $p = 9$, $t_{\rm fit}v = 1.0$ for temperature $T/v = 0.5$ 
  at on-site interaction strengths $U/v \in \{2.0,4.0,6.0,8.0\}$.
}
\label{fig:lp_Aw_diffU}
\end{figure}
Using linear prediction, we revisit two aspects of  single-site dynamical mean field theory: the opening of the Mott gap as interaction strength is increased, and the establishment of a quasiparticle peak as temperature is lowered in the coherent metallic regime.

Fig.~\ref{fig:lp_Aw_diffU} shows the data of Fig.~\ref{fig:DMFT_Aw_diffU} obtained within linear prediction. It is evident that the increased frequency resolution leads to additional features in the spectral function. $U/v=2$ is metallic with little change of the band structure due to correlations. $U/v=4$ shows `bad metallic' behavior where the spectral function near zero is suppressed due to the onset of insulating correlations. As the interaction is raised to $U/v=6$, a gap opens and a double-peak feature develops, and by $U/v=8$ a clear insulating structure with a wide Hubbard gap has developed.

\begin{figure}[htb]
\includegraphics[width=\columnwidth]{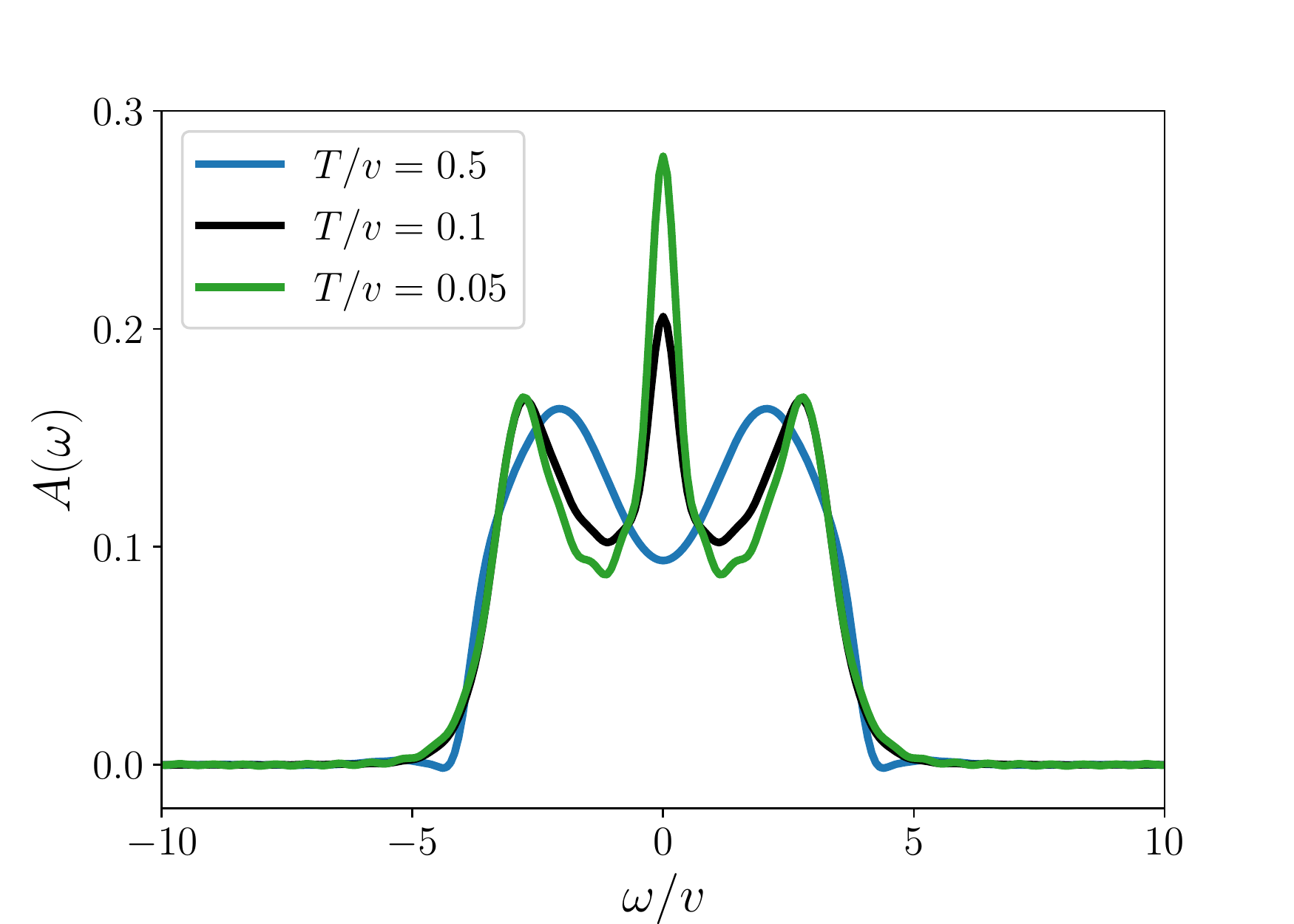}
\caption{
  The converged DMFT spectral function $A(\omega)$ obtained by extrapolating
  the real-time Green's function from $tv = 2.0$ to $tv = 10.0$ using linear
  prediction with $p = 9$, $t_{\rm fit}v = 1.0$ for $U/v = 4$
  at temperatures $T/v \in \{0.5, 0.1, 0.05\}$. 
}
\label{fig:lp_Aw_diffT}
\end{figure}
As temperature is lowered at $U/v=4$, the `bad metal' dip in the spectral 
function disappears, and for a temperature of $T/v=0.1$, a clear sign of a 
`quasiparticle peak' has developed (Fig.~\ref{fig:lp_Aw_diffT}). 
Further lowering of the temperature makes the quasiparticle peak more 
pronounced, while deepening the `dip' between the Hubbard side bands and the peak.

\section{\label{sec:conclusions}Conclusions}
We have presented a first application of real-time quantum Monte Carlo methods 
to real-time dynamical mean field theory. We showed that it is possible to 
obtain accurate Green's functions and self-energies directly in real-time. We 
further showed that if one is interested in frequency-transformed properties, 
linear prediction methods for extending the range of the available data to 
longer times work in practice for Monte Carlo data with sufficient accuracy.

We expect that in the near future our algorithms will have applications mainly out of equilibrium, in the study of quenches and driven problems, which the impurity solver algorithm demonstrated here can simulate at no additional cost. In contrast, reaching substantially longer times or substantially lower temperatures than the ones demonstrated here will require additional optimizations and access to supercomputer resources.

To take a long view on the equilibrium problem, it is useful to consider the respective scaling properties of imaginary-time and real-time algorithms with regard to the desired accuracy of real frequency quantities. Whereas imaginary-time algorithms are currently far more efficient, they are limited by the exponential sensitivity to errors in the various analytical continuation methods. Real-time algorithms are substantially more expensive, but their accuracy is limited only by the maximum simulated time. In the inchworm method, extending this time requires a quadratic increase in computational effort. It is therefore entirely conceivable that in the future---as the field grows to be interested in larger, more frustrated problems and in higher frequency resolution---the real-time algorithms will eventually surpass their imaginary-time counterparts in efficiency by the sheer power of scaling.

\section{\label{sec:acknowledgments}Acknowledgments}
Funding for this project was provided by DOE ER 46932. This
research used resources of the National Energy Research
Scientific  Computing  Center,  a  DOE  Office  of  Science
User  Facility  supported  by  the  Office  of  Science  of  the
U.S.  Department  of  Energy  under  Contract  No. DE-AC02-05CH11231. GC  was  supported  by  the  ISRAEL SCIENCE FOUNDATION (grant No. 1604/16)

\appendix
\begin{figure}[htb]
 	\includegraphics[width=\columnwidth]{{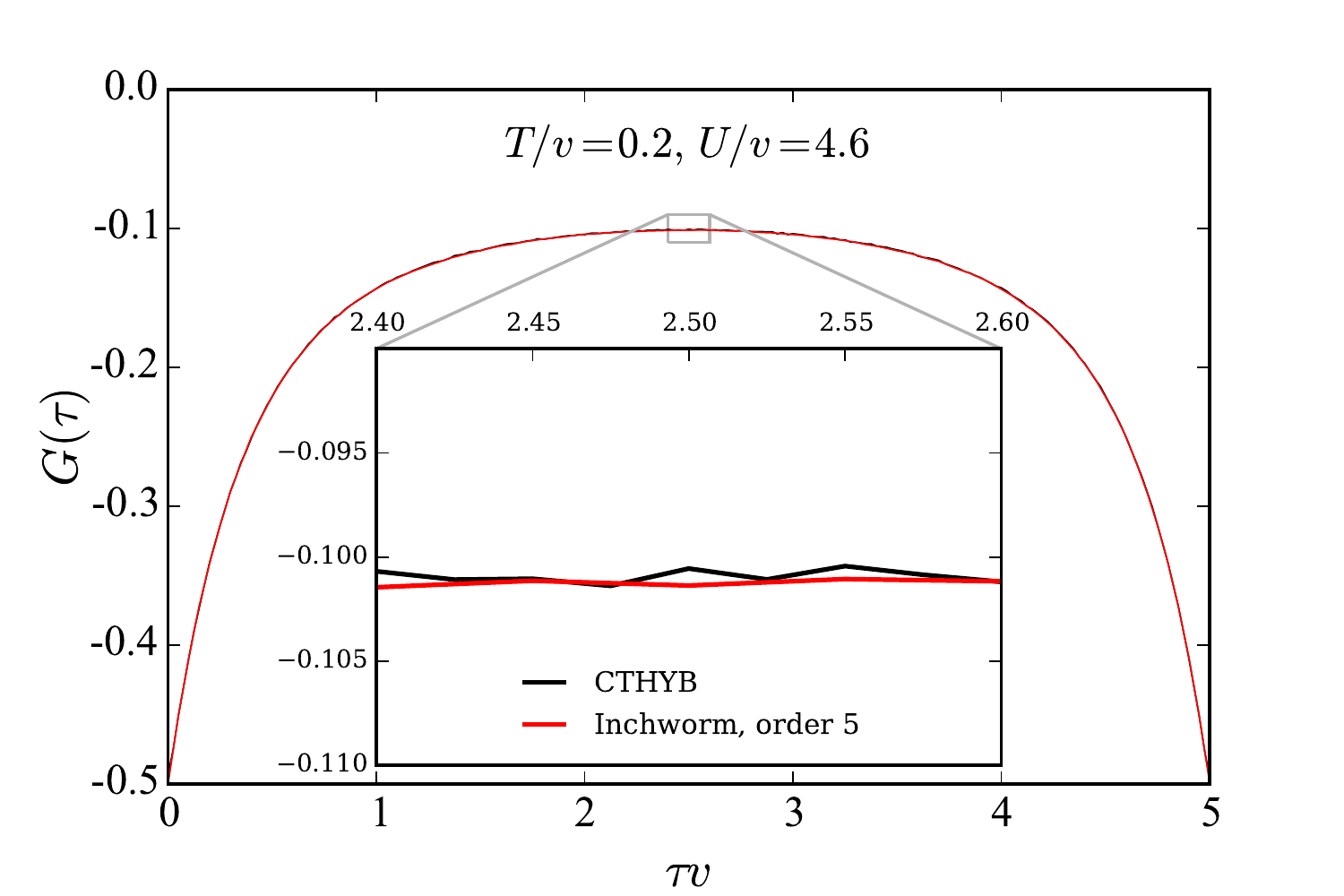}}
	\includegraphics[width=\columnwidth]{{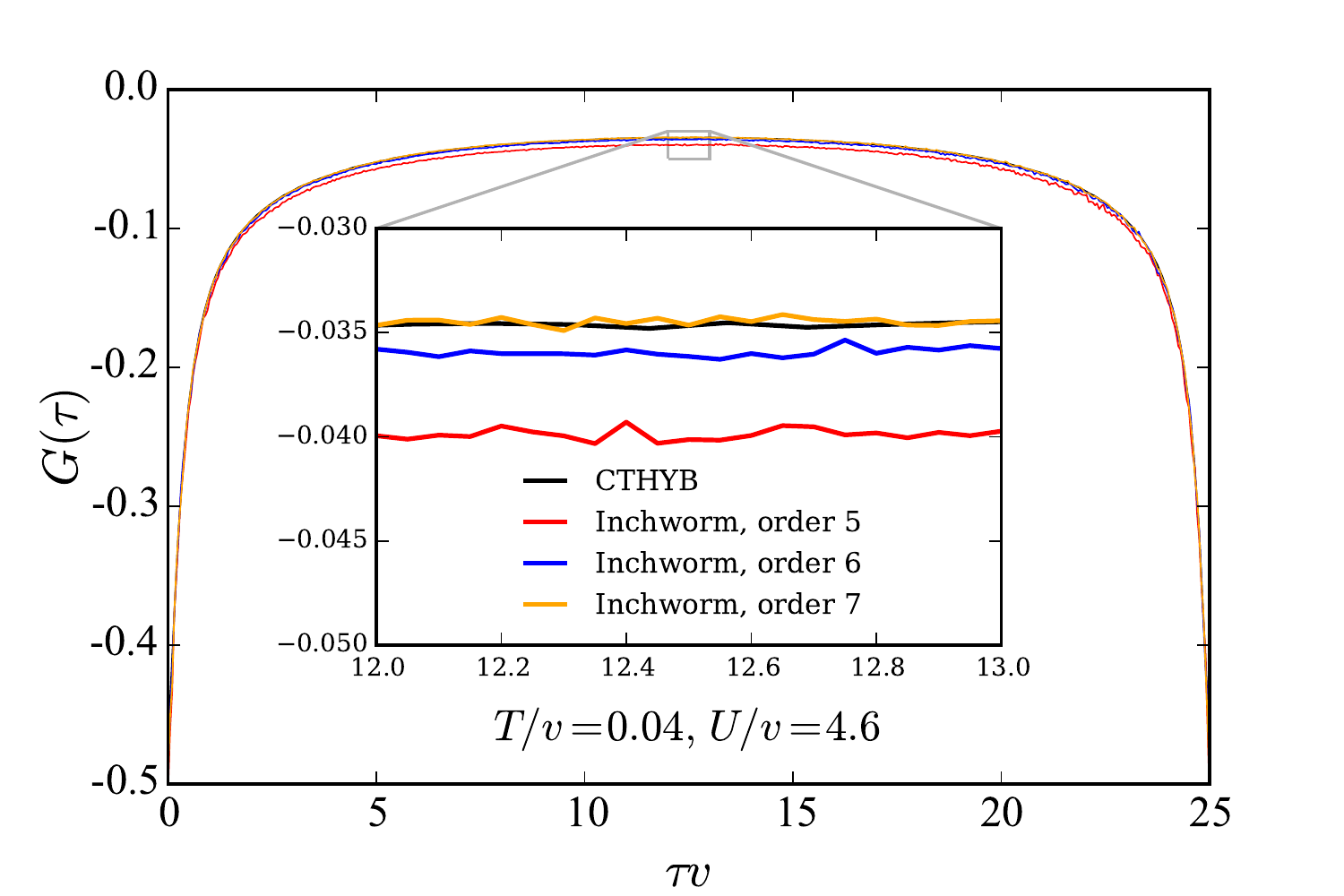}}
    \label{fig:Gtau_comp_with_eq_beta25}
    \caption{
		Matsubara Green's function $G(\tau)$ computed for the impurity model in 
		equilibrium with $U/v = 4.6$ at two temperatures and with different inchworm
        order truncations. Results from an equilibrium hybridization expansion solver
        (TRIQS/CTHYB\cite{TRIQS,CTHYB}) are given as a reference.
    }
    \label{fig:Gtau_comp_with_eq}
\end{figure}
\section{\label{sec:inchworm_order}Convergence with respect to the inchworm 
truncation order}
Fig.~\ref{fig:Gtau_comp_with_eq} shows the comparison of the Matsubara Green's function $G(\tau)$ computed by the inchworm nonequilibrium QMC solver and an equilibrium continuous-time hybridization expansion solver\cite{Werner06} (we used the solver of the open source TRIQS library\cite{TRIQS,CTHYB}). The data are obtained for an impurity calculation starting with a semielliptic density of states with a full bandwidth of $4$ at $U/v = 4.6$ and half-filling. The upper panel of Fig.~\ref{fig:Gtau_comp_with_eq} demonstrates that at high temperature $T/v = 0.2$, a maximum order of $5$ in the inchworm order truncation is sufficient. In contrast, as the temperature is lowered to $T/v = 0.04$ (near the onset of the first order coexistence in the dynamical mean field solution of this model), convergence of the inchworm calculations to the equilibrium result is only achieved at orders $5,6$ and $7$. The correspondence between inchworm order truncation and the number of crossings in an N-crossing approximation (such as NCA, OCA, or the two-crossing approximation) implies that a large number of crossings is essential for obtaining correct results in the correlated metallic regime.

\section{\label{app:linpred_robust}Robustness of linear prediction}
\begin{figure}[htb]
\includegraphics[width=\columnwidth]{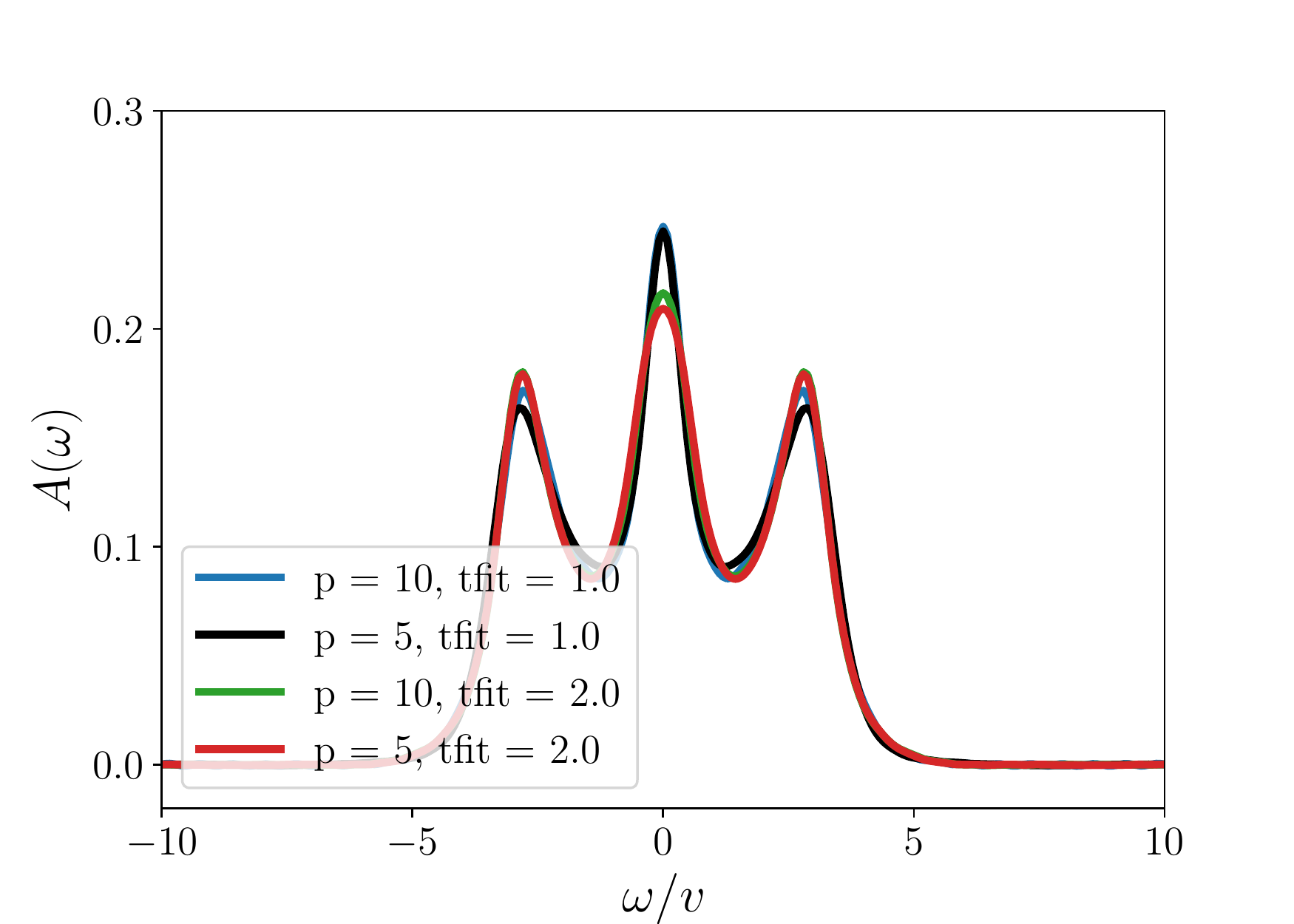}
\caption{
  The converged DMFT spectral function $A(\omega)$ obtained by extrapolating
  the real-time Green's function from $t = 2.0$ to $t = 10.0$ using linear
  prediction with $p = \{5, 10\}$, $t_{\rm fit} = \{1.0, 2.0\}$ for $U = 4$ at temperatures $T = 0.05$.
}
\label{fig:lp_Aw_metal_diffparams}
\end{figure}
\begin{figure}[htb]
\includegraphics[width=\columnwidth]{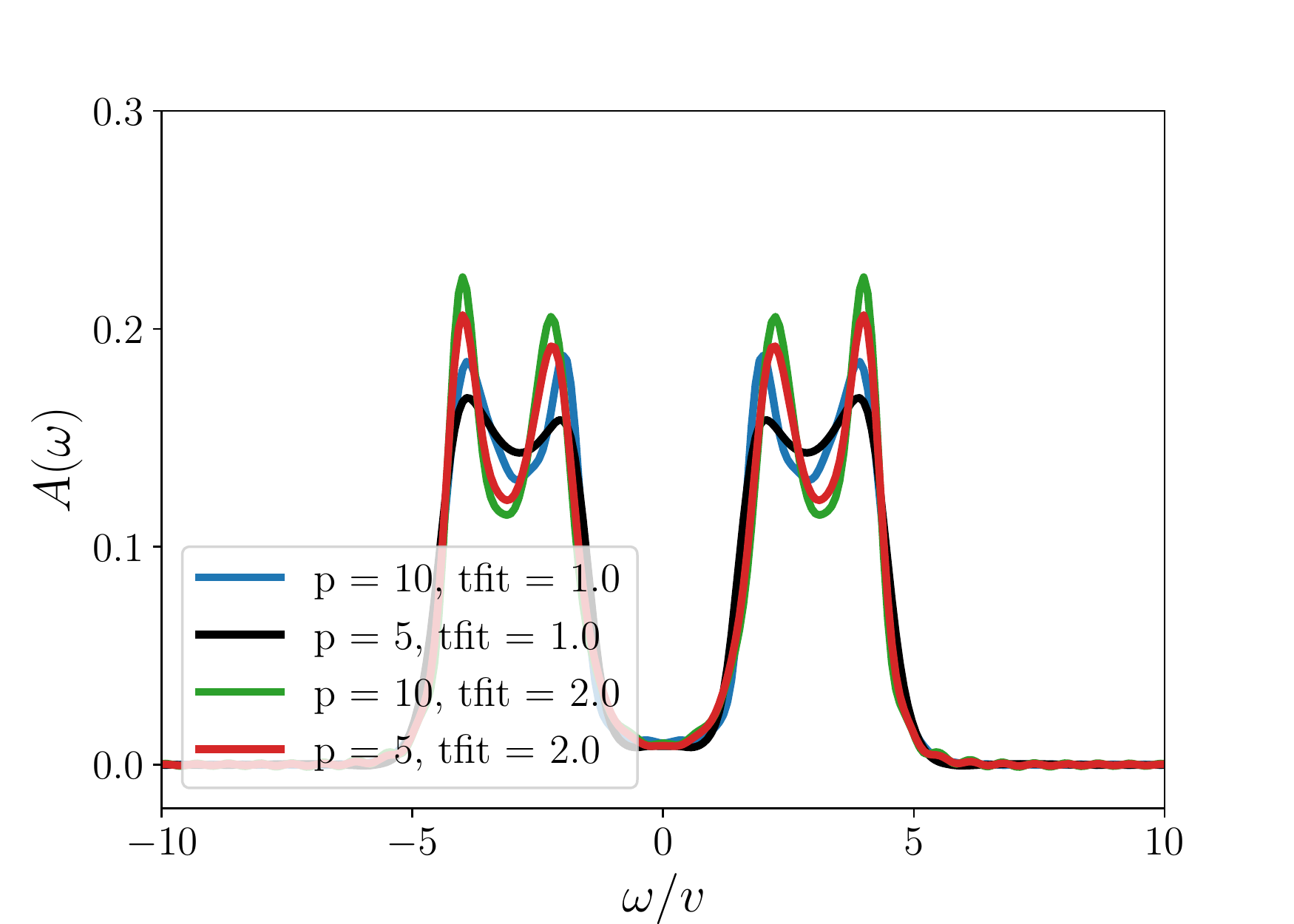}
\caption{
  The converged DMFT spectral function $A(\omega)$ obtained by extrapolating
  the real-time Green's function from $tv = 2.0$ to $tv = 10.0$ using linear
  prediction with $p = \{5, 10\}$, $t_{\rm fit}v = \{1.0, 2.0\}$ for $U/v = 6$
  at temperature $T/v = 0.5$.
}
\label{fig:lp_Aw_insulator_diffparams}
\end{figure}
Linear prediction, introduced in Sec.~\ref{sec:linpred} in the main text, is a method used to extrapolate data known up to a finite cutoff time to much longer times, such that smooth spectral functions can be extracted. The method can only succeed if the underlying data contains enough information to represent the long-time behavior accurately, and is expected to fail if this is not the case. In practice, two control parameters are available. First, the number of components (poles) $p$ that is being fitted. Second, the time interval over which the data is extrapolated.

Fig.~\ref{fig:gf_lp_validation} and Fig.~\ref{fig:spectrum_lp_validation} in the main text show the behavior of the spectral function and the real-time Green's function on the final time. Fig.~\ref{fig:lp_Aw_metal_diffparams} and Fig.~\ref{fig:lp_Aw_insulator_diffparams} show the dependence of the converged DMFT spectral functions on the number of poles $p$ and the maximum fitting time. Plotted are the converged dynamical mean field spectral functions obtained by extrapolating the real-time Green's function from $tv = 2$ to $tv = 10$.

Clear differences in the data are visible in the height of the quasi-particle peak, the height of the four peaks in the insulating regime, and the size of the dip separating the quasiparticle feature from the Hubbard side bands. The ambiguities become less pronounced when the time accessed is increased. In any case, the main features, in particular the existence of a quasiparticle peak in the metal or of a double peak structure above and below the Mott gap are clearly evident in our data.

\section{\label{app:sigma_technical}Self-energy extraction: technical issues}

\begin{figure*}[h]
 	\includegraphics[width=\columnwidth]{{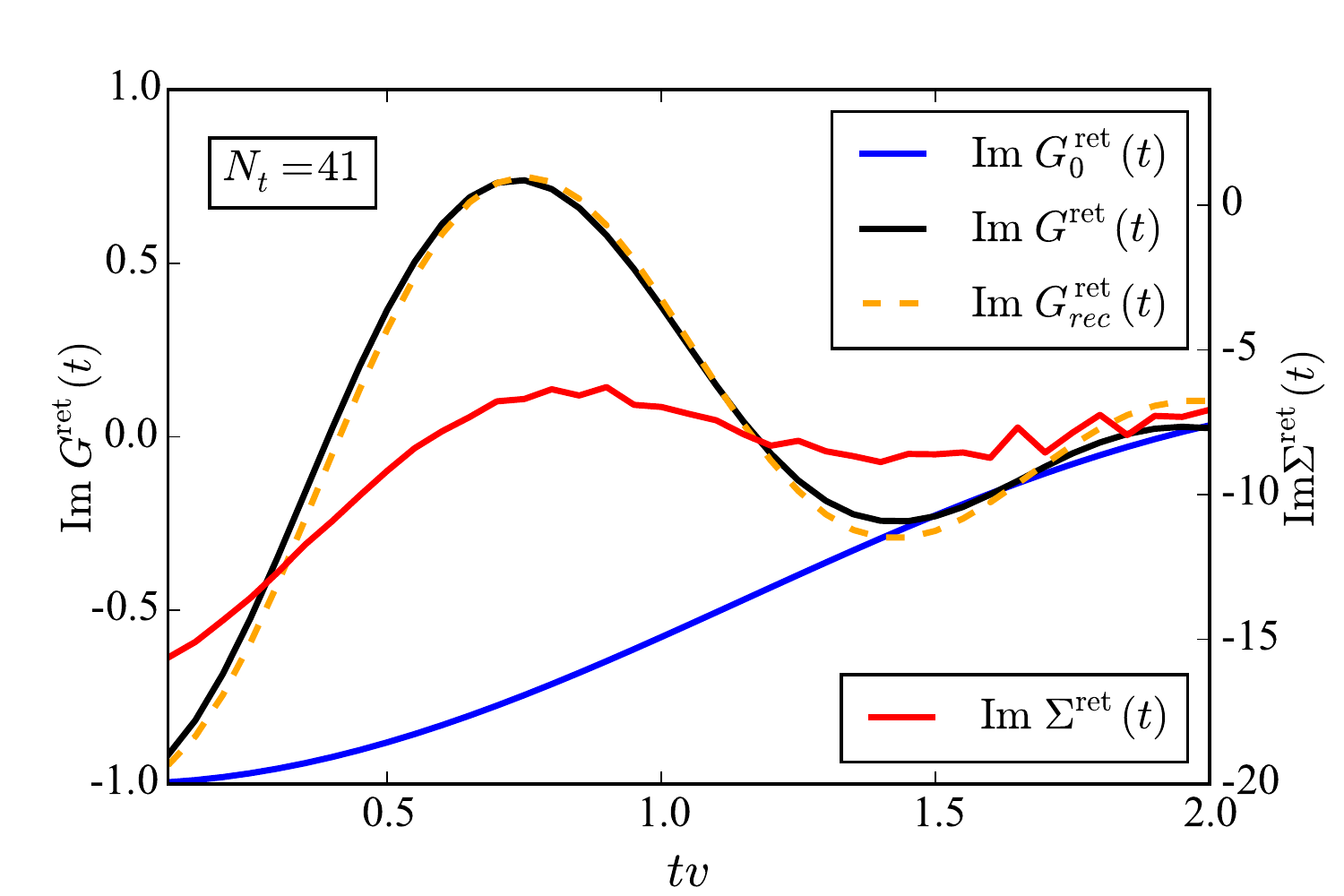}}
	\includegraphics[width=\columnwidth]{{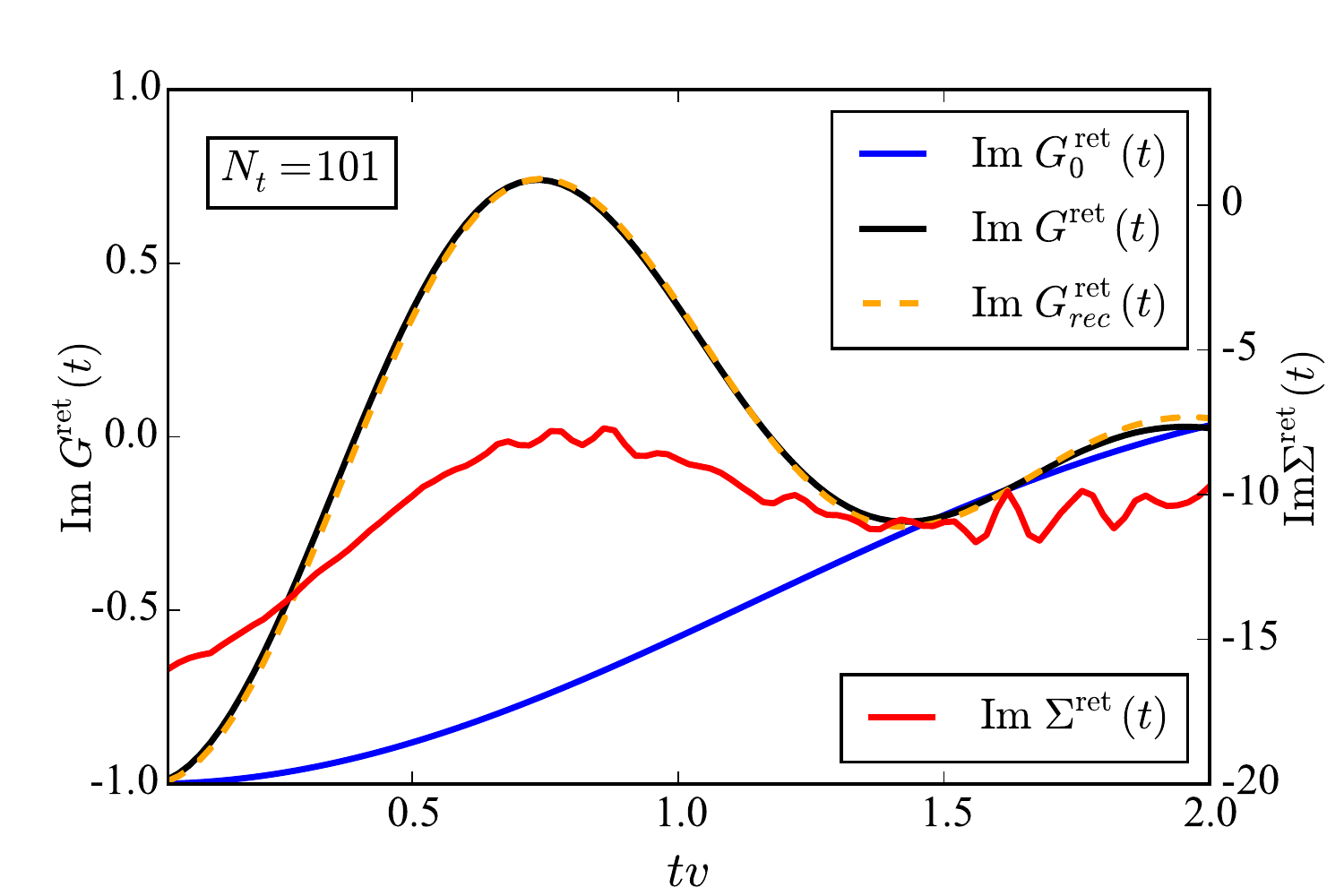}}
 	\includegraphics[width=\columnwidth]{{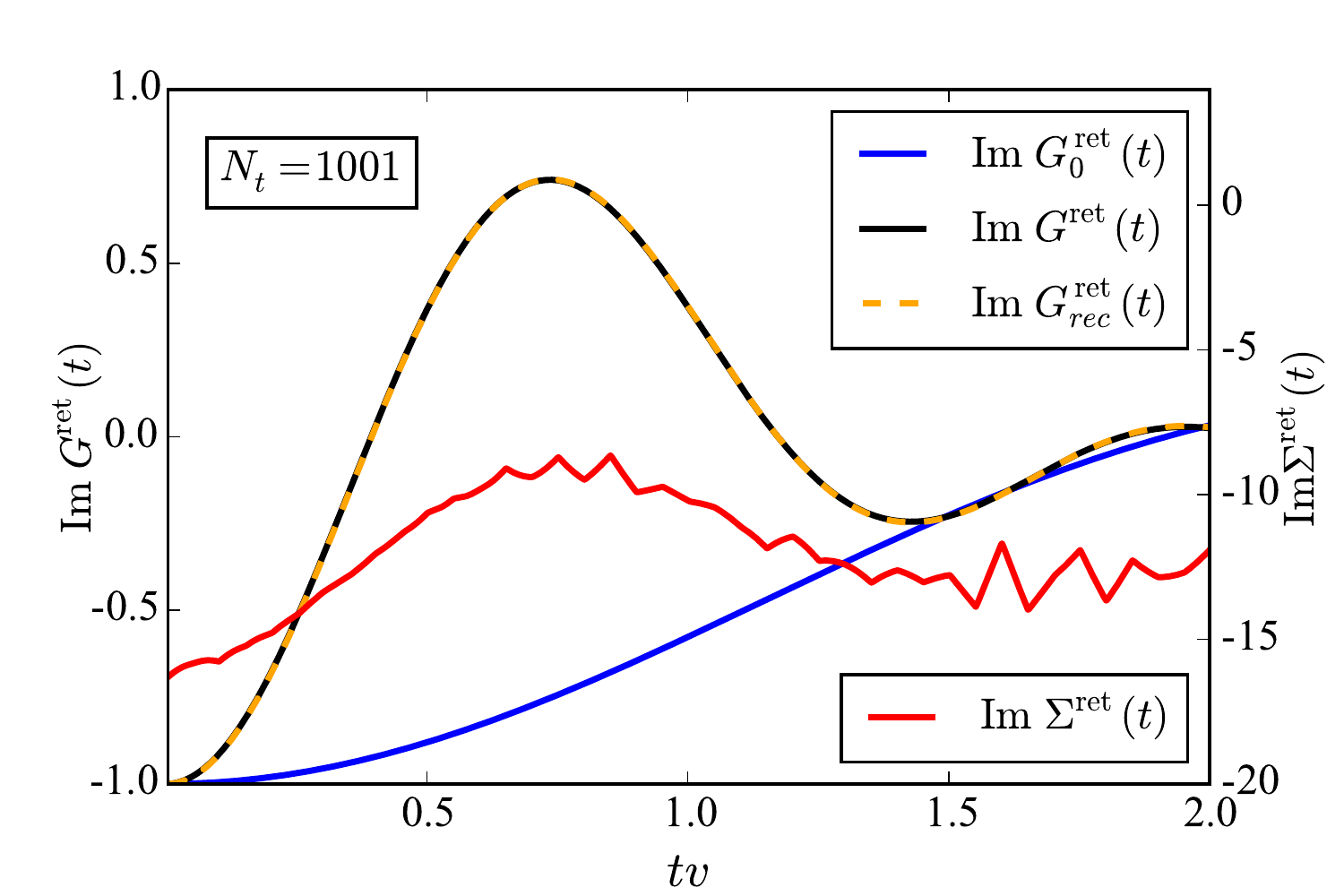}}
 	\includegraphics[width=\columnwidth]{{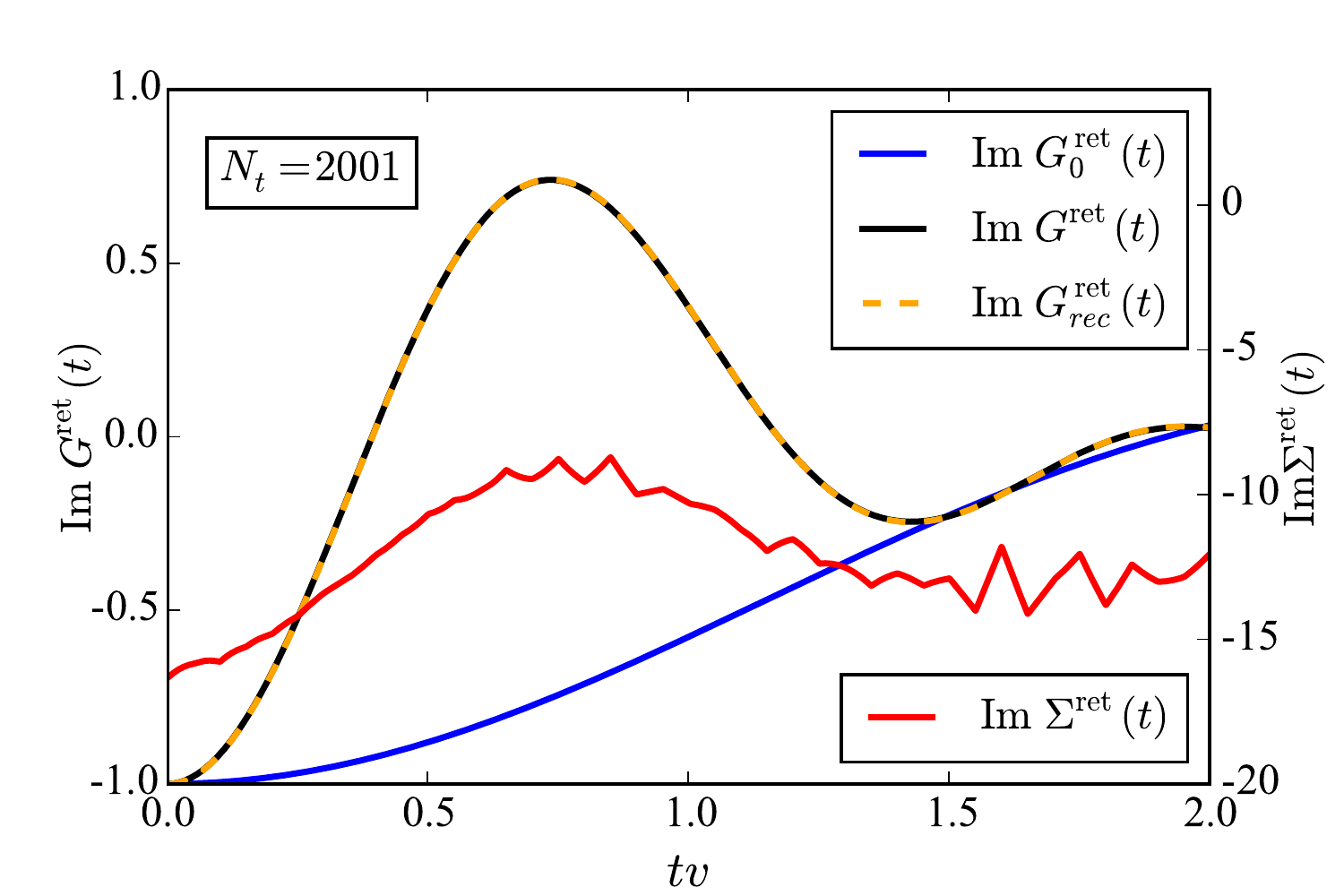}}
    \caption{
		Retarded components of the DMFT Green's function, bare Green's function and
        self-energy computed in equilibrium with $U/v = 8.0$ at $T/v = 0.5$.
        The self-energy curves are obtained by a direct solution of the Dyson equation
        in its discretized matrix form.
        $G^\mathrm{ret}(t)$ has been measured on 41 time slices, while a larger number of slices
        and cubic interpolation have been used to perform matrix inversions. The four
        subplots correspond to different numbers of interpolation points.
        $G_{rec}^\mathrm{ret}(t)$ (orange curve) is a Green's function reconstructed by iterative
        substitution of $\Sigma^\mathrm{ret}(t)$ into the Dyson equation.
        Top left: 41 slices. Top right: 101 slices. Bottom left: 1001 slices. Bottom 
        right: 2001 slices.
        }
    \label{fig:Sigma_interpol}
\end{figure*}

Being limited by the complexity of the inchworm algorithm, we have to take a 
sparse time grid for the real-time  Green's function $G_{\sigma}(t)$ and its retarded
counterpart $G^\mathrm{ret}_{\sigma}(t)$. The grid step $\Delta t v$ used throughout this work
is equal to 0.05, while $t_\mathrm{max}v\in\{2.0, 4.0\}$.

The self-energy extraction procedure presented in section \ref{sec:sigma} involves a
rectangular rule discretization of the time integrals, a first order scheme in
$\Delta t$. We have found that effect of the discretization error rapidly grows
as we propagate to larger times in $\Sigma^\mathrm{ret}_{\sigma}(t)$. It is
therefore crucial to introduce a finer time grid, and interpolate $G^\mathrm{ret}_{\sigma}(t)$/
$G^\mathrm{ret}_{0,\sigma}(t)$ between the original nodes before doing matrix inversions.
Fig. \ref{fig:Sigma_interpol} shows self-energy extraction results for $T/v=0.5,\ U/v=8.0$,
computed with $N_t = 41$ (the original number of points), $101$, $1001$ and $2001$ interpolation
slices. One can clearly see a drastic difference in the self-energy curves for $N_t = 41$
and $2001$. On the other hand, there is no visible difference between $N_t = 1001$ and $2001$, which means the interpolation has converged.

The dashed orange curves are obtained by back-substitution of $\Sigma^\mathrm{ret}(t)$
into a trapezoidal-rule discretization of the Dyson equation (\ref{dyson})
(different choice of $w_{ij}$). For $N_t = 1001, 2001$ they lie on top of the input
Green's function, which also signals convergence.

\section{\label{app:short_time}Short time limit of $\Sigma^\mathrm{ret}(t)$}

\begin{equation}
	\Sigma^\mathrm{ret}(t) = \int_{-\infty}^{+\infty} \frac{d\omega}{2\pi}
		e^{-i\omega t} \Sigma^\mathrm{ret}(\omega).
\end{equation}
Let us introduce a rescaled frequency $z = \omega t$ and consider a
short time limit of the self-energy,
\begin{equation}
	\Sigma^\mathrm{ret}(0^+) = \lim_{t\to0^+} \frac{1}{t}
    	\int_{-\infty}^{+\infty} \frac{dz}{2\pi} e^{-iz} \Sigma^\mathrm{ret}(z/t).
\end{equation}

$\Sigma^\mathrm{ret}(z/t)$ is an analytic function in the upper half-plane of $z$
for any positive $t$. We now employ a high-frequency expansion of the
self-energy in the absence of a Hartree-Fock term,
\begin{equation}
	\Sigma^\mathrm{ret}(\omega) = \lim_{\delta\to0^+}\sum_{m=1}^\infty
    				   \frac{C^{(m)}}{(\omega+i\delta)^m},
\end{equation}
\begin{equation}
	\Sigma^\mathrm{ret}(0^+) = \lim_{t\to0^+} \frac{1}{t}
    	\lim_{\delta\to0^+}\sum_{m=1}^\infty
    	\int_{-\infty}^{+\infty} \frac{dz}{2\pi} e^{-iz}
        \frac{t^m C^{(m)}}{(z+i\delta t)^m}
\end{equation}

It is easy to see that only the $m=1$ term contributes in the limit of $t\to0^+$.
\begin{multline}
	\Sigma^\mathrm{ret}(0^+) = \lim_{t\to0^+} \lim_{\delta\to0^+}
    	\int_{-\infty}^{+\infty} \frac{dz}{2\pi} e^{-iz}
        \frac{C^{(1)}}{z+i\delta t} =\\=
    \lim_{t\to0^+} \lim_{\delta\to0^+} (-ie^{-\delta t} C^{(1)}) = -iC^{(1)}.
\end{multline}

For the symmetric single orbital Anderson model $C^{(1)} = U^2/4$ (independent of bath parameters).
For a derivation see, for instance, Ref.~\onlinecite{Potthoff1997}. Given this,
\begin{equation}
	\Sigma^\mathrm{ret}(0^+) = -iU^2/4.
\end{equation}
\cleardoublepage

\bibliographystyle{apsrev4-1}
\bibliography{inchworm_dmft}

\end{document}